\documentclass
	[
		10 pt
	]
	{article}
\usepackage{tikz, pgf, pgfplots}
\usetikzlibrary
	{
		3d,
		arrows,											
		arrows.meta,
		automata,										
		backgrounds,
		bending,
		calc,
		chains,
		datavisualization.formats.functions,			
		decorations.pathmorphing,
		decorations.pathreplacing,
		decorations.text,
		fadings,
		fit,
		graphs,
		graphs.standard,
		matrix,
		positioning,									
		quantikz2,
		quotes,
		shadings,
		shadows.blur,
		shapes,
		trees
	}
\pgfplotsset{compat = newest}
\usepgflibrary
	{
		shapes.misc
	}
\usepgfplotslibrary
	{
		dateplot
	}
\usepackage{tikzpeople}
\usepackage{fontawesome5}

\usepackage{amsthm}
\usepackage{amssymb}
\usepackage[bb = boondox]{mathalfa}
\usepackage{dsfont}
\usepackage{newtxmath}
\usepackage{mathtools}
\usepackage{multicol}
\usepackage{braket}
\usepackage{physics}
\usepackage{empheq}

\usepackage[ compat = newest ]{yquant}

\usepackage{xcolor}
\usepackage{varwidth}
\usepackage[ skins, breakable, listings, theorems ] {tcolorbox}
\usepackage{graphicx}
\usepackage{xurl}                                                                                  %
\usepackage{hyperref}
\hypersetup
	{
		colorlinks,
		linkcolor = {WordRed!75!black},
		citecolor = {WordBlueDarker25},
		urlcolor = {WordAquaDarker50}
	}
\usepackage{colortbl}
\usepackage{float}
\usepackage{nicematrix}
\usepackage{cellspace}
\usepackage{makecell}
\usepackage{multirow}
\usepackage[ linesnumbered, tworuled, vlined ] {algorithm2e}
\usepackage{appendix}
\usepackage{enumitem}
\usepackage{siunitx}
\usepackage{xintexpr}
\usepackage{spreadtab}
\usepackage{framed}
\usepackage{svg}
\usepackage{lipsum}
\newcounter{mathseed}
\pgfmathsetseed{\arabic{mathseed}}
\pgfdeclarelayer{background}
\pgfsetlayers{background, main}
\pgfdeclaredecoration{irregular fractal line}{init}
{
	\state{init}[width=\pgfdecoratedinputsegmentremainingdistance]
	{
		\pgfpathlineto{\pgfpoint{random*\pgfdecoratedinputsegmentremainingdistance}{(random*\pgfdecorationsegmentamplitude-0.02)*\pgfdecoratedinputsegmentremainingdistance}}
		\pgfpathlineto{\pgfpoint{\pgfdecoratedinputsegmentremainingdistance}{0pt}}
	}
}
\def\tornpaper#1{%
	\ifthenelse{\isodd{\value{mathseed}}}
	{%
		\tikz
		{
			\node[inner sep = 1em] (A) {#1};		
			\begin{pgfonlayer}{background}			
				\fill[paper]						
				\pgfextra{\pgfmathsetseed{\arabic{mathseed}}\addtocounter{mathseed}{1}}%
				{decorate[irregular cloudy border]{decorate{decorate{decorate{decorate[ragged border]{
										(A.north west) -- (A.north east)
				}}}}}}
				-- (A.south east)
				\pgfextra{\pgfmathsetseed{\arabic{mathseed}}}%
				{decorate[irregular spiky border]{decorate{decorate{decorate{decorate[ragged border]{
										-- (A.south west)
				}}}}}}
				-- (A.north west);
			\end{pgfonlayer}
		}
	}
	{%
		\tikz{
			\node[inner sep=1em] (A) {#1};  
			\begin{pgfonlayer}{background}  
				\fill[paper] 
				\pgfextra{\pgfmathsetseed{\arabic{mathseed}}\addtocounter{mathseed}{1}}%
				{decorate[irregular spiky border]{decorate{decorate{decorate{decorate[ragged border]{
										(A.north east) -- (A.north west)
				}}}}}}
				-- (A.south west)
				\pgfextra{\pgfmathsetseed{\arabic{mathseed}}}%
				{decorate[irregular cloudy border]{decorate{decorate{decorate{decorate[ragged border]{
										-- (A.south east)
				}}}}}}
				-- (A.north east);
		\end{pgfonlayer}}
	}
}
\definecolor{MyLightRed}{RGB}{244, 213, 245}
\definecolor{WordRed}{RGB}{255, 0, 102}
\definecolor{WordRedAccent5Lighter60}{HTML}{F5B5A7}
\definecolor{WordRedAccent5Darker25}{HTML}{B23214}
\definecolor{RedDarkLightest}{HTML}{ff0088}
\definecolor{RedDarkLight}{HTML}{ea005f}
\definecolor{RedPurple}{HTML}{aa007f}
\definecolor{Purple}{HTML}{911146}
\definecolor{PurpleDark}{RGB}{102, 0, 102}
\definecolor{WordLightGreen}{RGB}{140, 214, 192}
\definecolor{WordGreen}{RGB}{0, 176, 80}
\definecolor{GreenLightest}{HTML}{00ffa0}
\definecolor{GreenLighter1}{HTML}{00b383}
\definecolor{GreenLighter2}{HTML}{00aa7f}
\definecolor{GreenDark}{HTML}{225522}
\definecolor{GreenTeal}{HTML}{008080}
\definecolor{WordIceBlue}{RGB}{223, 227, 229}
\definecolor{MyVeryLightBlue}{RGB}{211, 245, 247}
\definecolor{WordBlueVeryLight}{RGB}{0, 176, 240}
\definecolor{WordBlueLight}{RGB}{0, 112, 192}
\definecolor{WordBlueDark}{RGB}{46, 116, 181}
\definecolor{WordBlueDarker}{RGB}{31, 78, 121}
\definecolor{WordBlueDarker25}{RGB}{54, 96, 146}
\definecolor{WordBlueDarker50}{RGB}{36, 64, 98}
\definecolor{WordBlueDarkest}{RGB}{0, 32, 96}
\definecolor{WordBlue}{RGB}{19, 65, 99}
\definecolor{MyBlue}{RGB}{0, 64, 128}
\definecolor{MyDarkBlue}{RGB}{0, 51, 102}
\definecolor{BlueVeryDark}{HTML}{222255}
\definecolor{MagentaVeryLight}{RGB}{178, 162, 201}
\definecolor{MagentaLighter}{RGB}{161, 106, 221}
\definecolor{MagentaLight}{RGB}{128, 100, 162}
\definecolor{MagentaDark}{RGB}{106, 65, 152}
\definecolor{MagentaVeryDark}{RGB}{97, 75, 128}
\definecolor{WordAquaLighter80}{RGB}{218, 238, 243}
\definecolor{WordAquaLighter60}{RGB}{183, 222, 232}
\definecolor{WordAquaLighter40}{RGB}{146, 205, 220}
\definecolor{WordAquaDarker25}{RGB}{49, 134, 155}
\definecolor{WordAquaAccent2Darker25}{HTML}{398E98}
\definecolor{WordAquaDarker50}{RGB}{33, 89, 103}
\definecolor{WordVeryLightTeal}{RGB}{223, 236, 235}
\definecolor{WordLightTeal}{RGB}{160, 199, 197}
\definecolor{WordDarkTealLighter80}{RGB}{207, 223, 234}
\definecolor{WordDarkTeal}{RGB}{72, 123, 119}
\definecolor{WordDarkerTeal}{RGB}{48, 82, 80}
\definecolor{WordTurquoiseLighter80}{RGB}{209, 238, 249}
\definecolor{WordGoldAccent1Lighter40}{HTML}{FFDF6A}
\definecolor{WordGoldAccent1Darker25}{HTML}{C49A00}
\definecolor{Brown}{HTML}{666633}
\definecolor{WordOrangeAccent2Lighter60}{HTML}{FCD3A4}
\definecolor{WordOrangeAccent4Lighter60}{HTML}{F7C5A1}

\newtheorem{definition}{Definition}[section]

\usepackage
	[
		a4paper,
		left = 2.5cm,
		right = 2.5 cm,
		top = 2.5cm,
		bottom = 3.0 cm
	]
	{geometry}
\title
	{
		A distributed and parallel $(k, n)$ QSS scheme with verification capability
	}

\usepackage{orcidlink}
\newcommand{\orcidicon}[1]{\href{https://orcid.org/#1}{\includegraphics[height=\fontcharht\font`\B]{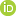}}}

\author
{
	Theodore Andronikos
	\orcidicon{0000-0002-3741-1271}
	\\
	Department of Informatics, Ionian University, \\
	7 Tsirigoti Square, 49100 Corfu, Greece; \\
	andronikos@ionio.gr \\
}
\begin{document}

\maketitle

\begin{abstract}
	This article introduces a novel Quantum Secret Sharing scheme with $( k, n )$ threshold and endowed with verification capability. The new protocol exploits the power of entanglement and evolves in three phases. The primary novelty of the new protocol lies in its ability to operate completely parallelly in a fully distributed setup, where the spymaster and her agents all are in different locations, in contrast to the vast majority of analogous protocols that assume a localized scenario in which all agents are at the same place. The spymaster sends all necessary information to all intended recipients simultaneously in one step. All phases are executed in parallel, minimizing the overall execution cost of the protocol. Given its comparative complexity, we provide a comprehensive and detailed analysis to establish its information-theoretic security in the sense of preventing both outside eavesdroppers from obtaining any useful information and inside rogue agents from sabotaging its successful completion. The protocol eliminates the need for a quantum signature scheme or pre-shared keys, thereby simplifying the process and lowering complexity. Finally, the possibility of its implementation by contemporary quantum computers is promising because the protocol relies exclusively on CNOT and Hadamard gates and all players operating on similar or identical quantum circuits.
\end{abstract}
\section{Introduction} \label{sec: Introduction}

Our era witnesses a monumental quest to construct quantum computers with capabilities exceeding those of classical computers. Admittedly, this goal has not yet been fully realized. Nonetheless, significant advancements are evident, as illustrated by IBM's 127-qubit Eagle  \cite{IBMEagle}, 433-qubit Osprey \cite{IBMOsprey}, and the latest 1,121-qubit Condor \cite{IBMCondor}. These developments suggest a notable acceleration towards the quantum era of utility. It is obvious that the current state of quantum technology has advanced sufficiently to merit serious exploration in the design and implementation of cryptographic protocols. In this context, it is not surprising that quantum cryptographic techniques have gained substantial traction since the pioneering work of Bennett and Brassard \cite{Bennett1984}, which introduced quantum key distribution. Additionally, the influential quantum algorithms developed by Shor \cite{Shor1999} and Grover \cite{Grover1996} have highlighted the potential weaknesses of many prevalent security algorithms in the face of quantum computing. In the pursuit of cryptographic systems that remain secure against both quantum and classical threats, two primary strategies can be discerned. The first, often termed post-quantum or quantum-resistant cryptography, relies on problems that are presumed to be computationally challenging. The caveat here is that this assumption is more the result of accumulated experience, rather than a conclusive mathematical proof. The second, known as quantum cryptography, leverages the principles of quantum mechanics, harnessing the fundamental laws of nature to establish security. Currently, numerous active research domains can be identified, including quantum key distribution \cite{Bennett1984, Ekert1991, Bennett1992, Bennett2014} and the more recent \cite{Ampatzis2021}, quantum secret sharing \cite{Hillery1999, Cleve1999, Karlsson1999, Ampatzis2022}, quantum teleportation \cite{Bennett1993, Bouwmeester1997}, and quantum secure direct communication \cite{Deng2003, Deng2004, Wang2005}, among others.

In the classical domain, secret sharing was introduced in the landmark papers of Blakely \cite{Blakley1979} and Shamir \cite{Shamir1979}. One of the most prevalent symmetric methods to share a secret sharing is via a threshold scheme. A $( k, n )$ threshold scheme has $n$ shares, of which any $k$ are sufficient to reconstruct the secret, while any set of $k - 1$ or fewer convey no information about the secret. Blakely \cite{Blakley1979} and Shamir \cite{Shamir1979} showed that threshold schemes exist for all values of $k$ and $n$ with $n > k$. In particular, Shamir based his partitioning scheme on polynomial interpolation \cite{Shamir1979}. Later, in the groundbreaking works of Hillery et al. \cite{Hillery1999} and Cleve et al. \cite{Cleve1999}, it was established that secret sharing could be realized through quantum mechanics, thereby giving rise to the new field known as Quantum Secret Sharing (QSS from now on).

Since that time, significant advancements have been made in this area, with numerous proposals actively contributing to ongoing research \cite{Karlsson1999, smith2000quantum, gottesman2000theory, bandyopadhyay2000teleportation, xiao2004efficient, fortescue2012reducing, qin2020hierarchical, senthoor2022theory}. To verify the viability of QSS, several research groups have elaborate created protocols such as \cite{tittel2001experimental, bogdanski2008experimental, bell2014experimental}. Sophisticated proof-of-concept experimental demonstrations of novel and practical protocols appropriate for practical applications have been presented by researchers in recent years \cite{fu2015long, wu2020passive, grice2019quantum, gu2021secure}. Furthermore, a different line of research has attempted to expand the boundaries of the field by putting forth non-binary protocols that employ qudits rather than qubits \cite{tavakoli2015secret, pinnell2020experimental}. Some notable recent works include \cite{Gu2021, Yan2021, Li2022, Tian2023, Meng2023, Kuo2023}, and \cite{Zhang2024a}, one of the few involving spatially distributed players.

It is fairly common in the literature to view quantum protocols as quantum games. The famous Alice, Bob, and his clones, who serve as Alice's agents, as well as the infamous Eve, the external eavesdropper, all star in this particular game. The goal of using games in our presentation is to make technical concepts easier to understand. From their inception in 1999 \cite{Meyer1999, Eisert1999}, quantum games have become widely accepted. According to \cite{Andronikos2018, Andronikos2021, Andronikos2022a, Kastampolidou2023a}, quantum strategies may be better than classical ones, so this is not shocking. A notable example of this phenomenon is the well-known prisoners' dilemma game, which also applies to other abstract quantum games \cite{Eisert1999, Giannakis2019}. Recent works like \cite{Ampatzis2021, Ampatzis2022, Ampatzis2023, Andronikos2023, Andronikos2023a, Andronikos2023b, Karananou2024, Andronikos2024, Andronikos2024a} reveal that the presentation of protocols in the field of quantum cryptography frequently takes the form of games. Interestingly, political structures also use the quantization of classical systems \cite{Andronikos2022}. In the wider context of game-theoretic applications, unconventional settings, such as biological systems, have attracted considerable interest. It is noteworthy that biological systems may lead to biostrategies that surpass classical strategies, even in well-known games like the Prisoners' Dilemma \cite{Kastampolidou2020, Kastampolidou2021, Kastampolidou2023, Papalitsas2021, Adam2023}.

\textbf{Contribution}. This article presents a new Quantum Secret Sharing scheme with a $( k, n )$ threshold that is enhanced with verification capabilities. The innovative protocol harnesses the concept of entanglement and progresses through three distinct phases. A key highlight of the new protocol is its capacity to function entirely in parallel within a fully distributed environment, where the spymaster and her agents are situated in different geographical locations. This contrasts with most similar protocols that operate under the assumption that all agents are co-located. The spymaster transmits all necessary information to the intended recipients at once and completely in parallel, streamlining the process. During each phase the players' actions transpire concurrently, which helps to reduce the overall execution time of the protocol. Due to its complexity, we offer a thorough analysis to demonstrate its information-theoretic security, ensuring that neither external eavesdroppers can gain useful information nor internal rogue agents can disrupt its successful execution. The proposed protocol stands out because it does not depend on a quantum signature scheme or pre-shared keys, which simplifies its design and lowers its execution cost. Additionally, the implementation of the protocol promotes consistency, as all agents utilize the same private quantum circuits that closely resemble the spymaster's circuit. This approach creates a fully modular quantum system with identical modules. Each private quantum circuit uses only the commonly employed Hadamard and CNOT quantum gates, making it easy to implement on modern quantum computers.

\subsection*{Organization} \label{subsec: Organization}

This article is structured in the following way. Section \ref{sec: Introduction} introduces the topic and includes references to relevant literature. Section \ref{sec: Notation & Terminology} offers a brief overview of key concepts, which serves as a basis for grasping the technical details. In Section \ref{sec: An Overview of the DPVQSS Protocol}, a thorough explanation of the hypotheses that characterize the setting is provided. The protocol is formally introduced in Section \ref{sec: Detailed Analysis of the DPVQSS Protocol}. Section \ref{sec: Efficiency & Security Analysis} examines both the efficiency and security features of the protocol. Finally, the paper wraps up with a summary and a discussion of the protocol's nuances in Section \ref{sec: Discussion and Conclusions}.

\section{Notation \& terminology} \label{sec: Notation & Terminology}

\subsection{EPR pairs and GHZ states} \label{subsec: EPR Pairs and GHZ States}

One of the most renowned aspects of quantum mechanics is quantum entanglement, which forms the basis of most quantum protocols without exaggeration. One product state is insufficient to characterize entangled states of composite systems mathematically; instead, they must be expressed as a linear combination of two or more product states of their subsystems. The most basic type of maximal entanglement is represented by the famous Bell states, which are special quantum states of two qubits also referred to as EPR pairs. Below is the precise mathematical expression for each of the four Bell states. Qubits $\ket{ \cdot }_{ A }$ belong to Alice, and qubits $\ket{ \cdot }_{ B }$ belong to Bob. The subscripts $A$ and $B$ are used to highlight the subsystem to which the corresponding qubit belongs. Arguably, the most well-known examples of maximal entanglement are pairs of qubits in one of the four Bell states, also referred to as EPR pairs. 

\begin{tcolorbox}
	[
	grow to left by = 1.50 cm,
	grow to right by = 0.00 cm,
	colback = white,			
	enhanced jigsaw,			
	sharp corners,
	toprule = 0.1 pt,
	bottomrule = 0.1 pt,
	leftrule = 0.1 pt,
	rightrule = 0.1 pt,
	sharp corners,
	center title,
	fonttitle = \bfseries
	]
	
	\begin{minipage}[b]{0.475 \textwidth}
		\begin{align}
			\label{eq:Bell State Phi +}
			\ket{ \Phi^{ + } }
			=
			\frac { \ket{ 0 }_{ A } \ket{ 0 }_{ B } + \ket{ 1 }_{ A } \ket{ 1 }_{ B } }
			{ \sqrt{ 2 } }
		\end{align}
	\end{minipage} 
	\hfill
	\begin{minipage}[b]{0.45 \textwidth}
		\begin{align}
			\label{eq:Bell State Phi -}
			\ket{ \Phi^{ - } }
			=
			\frac { \ket{ 0 }_{ A } \ket{ 0 }_{ B } - \ket{ 1 }_{ A } \ket{ 1 }_{ B } }
			{ \sqrt{ 2 } }
		\end{align}
	\end{minipage}
	\begin{minipage}[b]{0.475 \textwidth}
		\begin{align}
			\label{eq:Bell State Psi +}
			\ket{ \Psi^{ + } }
			=\frac { \ket{ 0 }_{ A } \ket{ 1 }_{ B } + \ket{ 1 }_{ A } \ket{ 0 }_{ B } }
			{ \sqrt{ 2 } }
		\end{align}
	\end{minipage} 
	\hfill
	\begin{minipage}[b]{0.45 \textwidth}
		\begin{align}
			\label{eq:Bell State Psi -}
			\ket{ \Psi^{ - } }
			=
			\frac { \ket{ 0 }_{ A } \ket{ 1 }_{ B } - \ket{ 1 }_{ A } \ket{ 0 }_{ B } }
			{ \sqrt{ 2 } }
		\end{align}
	\end{minipage}
	
\end{tcolorbox}

In addition to EPR pairs, our protocol will also require a more general form of maximal entanglement. Maximal entanglement can be intuitively generalized for compound systems with $r$ qubits, where $r \geq 3$. For multipartite systems, the most well-known example of maximal entanglement is the $\ket{ GHZ_{ r } }$ state, where GHZ stands for the initials of the researchers Greenberger, Horne, and Zeilinger. Such a composite quantum system is made up of $r$ distinct qubits, each of which is thought of as a separate subsystem spatially separated from the other qubits. These $r$ qubits are all entangled in the $\ket{ GHZ_{ r } }$ state, which has the following mathematical description:

\begin{align}
	\label{eq: Extended General GHZ_n State}
	\ket{ GHZ_{ r } }
	=
	\frac
	{
		\ket{ 0 }_{ r - 1 } \ket{ 0 }_{ r - 2 } \dots \ket{ 0 }_{ 0 }
		+
		\ket{ 1 }_{ r - 1 } \ket{ 1 }_{ r - 2 } \dots \ket{ 1 }_{ 0 }
	}
	{ \sqrt{ 2 } }
	\ .
\end{align}

As with EPR pairs, to emphasize that $\ket{ GHZ_{ r } }$ entanglement requires $r$ distinct qubits, in equation \eqref{eq: Extended General GHZ_n State} we use the indices $i, \ 0 \leq i \leq r - 1$, to represent the $i^{ th }$ qubit. We adhere to this convention throughout this paper, and we write $\ket{ \cdot }_{ A }$, $\ket{ \cdot }_{ B }$, $\dots$, to indicate the qubits that belong to Alice, Bob, and so on. At the moment of writing this text, there exist quantum computers, such as the most recent IBM quantum computers \cite{IBMOsprey, IBMCondor}, that are able to prepare GHZ states by employing some of the most common quantum operations like the Hadamard and CNOT gates. Additionally, the ideal pathways used to generate these states are quite effective, as they only need $\lg r$ steps to produce the $\ket{ GHZ_{ r } }$ state ket \cite{Cruz2019}. Entanglement is a fascinating subject and we refer the interested reader to any relevant textbook, such as \cite{Nielsen2010, Yanofsky2013a, Wong2022}.

For the scope of the proposed protocol, a single EPR pair or a single $\ket{ GHZ_{ r } }$ tuple will not suffice. The mathematical description of the state of a compound system containing $p$ $\ket{ \Phi^{ + } }$ pairs, or $p$ $\ket{ GHZ_{ r } }$ tuples is shown below (see also \cite{Ampatzis2022} and \cite{Ampatzis2023}).

\begin{align}
	\label{eq: r-Fold Phi + Pairs}
	\ket{ \Phi^{ + } }^{ \otimes p }
	&=
	2^{ - \frac { p } { 2 } }
	\sum_{ \mathbf{ x } \in \mathbb{ B }^{ p } }
	\ket{ \mathbf{ x } }_{ A } \ket{ \mathbf{ x } }_{ B }
	\ ,
\end{align}

and

\begin{align}
	\label{eq: r-Fold Extended General GHZ_n State}
	\ket{ GHZ_{ r } }^{ \otimes p }
	&=
	2^{ - \frac { p } { 2 } }
	\sum_{ \mathbf{ x } \in \mathbb{ B }^{ p } }
	\ket{ \mathbf{ x } }_{ r - 1 } \dots \ket{ \mathbf{ x } }_{ 0 }
	\ .
\end{align}

In writing formulae \eqref{eq: r-Fold Phi + Pairs} and \eqref{eq: r-Fold Extended General GHZ_n State} the following notation is used.

\begin{itemize}
	\item	
	To avoid any ambiguity we make extensive use of subscripts to indicate which qubits belong to which subsystem.
	\item	
	$\mathbb{ B }$ is the binary set $\{ 0, 1 \}$.
	\item	
	To make clear when we refer to a bit vector $\mathbf{ x } \in \mathbb{ B }^{ p }$, we write $\mathbf{ x }$ in boldface, so as to distinguish from the single bit $x \in \mathbb{ B }$, written in regular typeface.
	\item	
	A bit vector $\mathbf{ x }$ is a sequence of $p$ bits: $\mathbf{ x } = x_{ p - 1 } \dots x_{ 0 }$. A special bit vector is the \emph{zero} bit vector, denoted by $\mathbf{ 0 }$, in which all the bits are zero, i.e., $\mathbf{ 0 } = 0 \dots 0$.
	\item	
	Each bit vector $\mathbf{ x } \in \mathbb{ B }^{ p }$ corresponds to one of the $2^{ p }$ basis kets that form the computational basis of the $2^{ p }$-dimensional Hilbert space.
\end{itemize}

\subsection{Inner product modulo $2$ operation} \label{subsec: Inner Product Modulo $2$ Operation}

The inner product modulo $2$ takes two bit vectors $\mathbf{ x }, \mathbf{ y } \in \mathbb{ B }^{ p }$, and returns their inner product $\mathbf{ x \bullet y }$. If $\mathbf{ x } = x_{ p - 1 } \dots x_{ 0 }$ and $\mathbf{ y } = y_{ p - 1 } \dots y_{ 0 }$, then $\mathbf{ x } \bullet \mathbf{ y }$ is defined as

\begin{align}
	\label{eq: Inner Product Modulo $2$}
	\mathbf{ x }
	\bullet
	\mathbf{ y }
	\coloneq
	x_{ p - 1 } y_{ p - 1 }
	\oplus \dots \oplus
	x_{ 0 } y_{ 0 }
	\ ,
\end{align}

where $\coloneq$ stands for ``is defined as,'' and $\oplus$ is \emph{addition modulo} $2$. An important formula that makes use of inner product modulo $2$ expresses the $p$-fold Hadamard transform of the basis ket $\ket{ \mathbf{ x } }$. Its proof can be found in most standard textbooks, e.g., \cite{Mermin2007, Nielsen2010}.

\begin{align}
	\label{eq: Hadamard p-Fold Ket x}
	H^{ \otimes p } \ket{ \mathbf{ x } }
	&=
	2^{ - \frac { p } { 2 } }
	\sum_{ \mathbf{ z } \in \mathbb{ B }^{ p } }
	( - 1 )^{ \mathbf{ z \bullet x } } \ket{ \mathbf{ z } }
	\ .
\end{align}

Our protocol relies on the following interesting property of the inner product modulo $2$ operation: for any fixed element $\mathbf{ c }$ of $\mathbb{ B }^{ p }$, other than $\mathbf{ 0 }$, it holds that for exactly half of the elements $\mathbf{ x } \in \mathbb{ B }^{ p }$, $\mathbf{ c } \bullet \mathbf{ x } = 0$, and for the remaining half $\mathbf{ c } \bullet \mathbf{ x } = 1$. Obviously, for the zero bit vector $\mathbf{ 0 }$ we have that for all $\mathbf{ x } \in \mathbb{ B }^{ p }$, $\mathbf{ c } \bullet \mathbf{ x } = 0$. Following \cite{Andronikos2023b}, we call this property the Characteristic Inner Product (CIP) property.

		\begin{align}
			\label{eq: Inner Product Modulo $2$ Property For Zero}
			\mathbf{ c } = \mathbf{ 0 }
			&\Rightarrow
			\text{for all } 2^{ p } \text{ bit vectors } \mathbf{ x } \in \mathbb{ B }^{ p },
			\text{ } \mathbf{ c } \bullet \mathbf{ x } = 0
			\\
			\label{eq: Inner Product Modulo $2$ Property For NonZero}
			\mathbf{ c } \neq \mathbf{ 0 }
			&\Rightarrow
			\left\{
			\
			\begin{matrix*}[l]
				\text{for } 2^{ p - 1 } \text{ bit vectors } \mathbf{ x } \in \mathbb{ B }^{ p }, \ \mathbf{ c } \bullet \mathbf{ x } = 0
				\\
				\text{for } 2^{ p - 1 } \text{ bit vectors } \mathbf{ x } \in \mathbb{ B }^{ p }, \ \mathbf{ c } \bullet \mathbf{ x } = 1
			\end{matrix*}
			\
			\right\}
		\end{align}

The proposed protocol also requires two other well-known states, $\ket{ + }$ and $\ket{ - }$, which are defined as

\begin{tcolorbox}
	[
	enhanced,
	breakable,
	grow to left by = 0.00 cm,
	grow to right by = 0.00 cm,
	colback = white,			
	enhanced jigsaw,			
	sharp corners,
	toprule = 0.1 pt,
	bottomrule = 0.1 pt,
	leftrule = 0.1 pt,
	rightrule = 0.1 pt,
	sharp corners,
	center title,
	fonttitle = \bfseries
	]
	\begin{minipage}[b]{0.45 \textwidth}
		\begin{align}
			\label{eq: Ket +}
			\ket{ + }
			=
			H \ket{ 0 } = \frac { \ket{ 0 } + \ket{ 1 } } { \sqrt{ 2 } }
		\end{align}
	\end{minipage} 
	\hfill
	\begin{minipage}[b]{0.45 \textwidth}
		\begin{align}
			\label{eq: Ket -}
			\ket{ - }
			=
			H \ket{ 1 } = \frac { \ket{ 0 } - \ket{ 1 } } { \sqrt{ 2 } }
		\end{align}
	\end{minipage}
\end{tcolorbox}

Decoys are typically prepared in these two states. All measurements during the execution of our protocol are performed with respect to the computational basis $\{ \ket{ 0 }, \ket{ 1 } \}$. However, for the purpose of eavesdropping detection, measurements are typically conducted with respect to the Hadamard basis.

\section{An overview of the DPVQSS protocol} \label{sec: An Overview of the DPVQSS Protocol}

We now proceed to describe in detail the proposed $( k, n )$ threshold Quantum Secret Sharing (QSS) scheme, with $k > \frac { n } { 2 }$. The main novelties of our protocol are its fully distributed and parallel traits. By this we mean that all actors, Alice and all her agents, are all spatially distributed in different locations. Moreover, the quantum and classical communications take place in a fully parallel manner. In addition, our protocol provides a method for verification that effectively acts as a fail-safe. The chosen name for this protocol is DPVQSS to emphasize its distributed and parallel traits, as well as its verification capability. From now on we shall simply refer to it by its acronym DPVQSS.

Let us first formalize the hypotheses underlying the operation of the DPVQSS protocol.

\begin{enumerate}
	[ left = 0.500 cm, labelsep = 1.000 cm, start = 1 ]
	\renewcommand \labelenumi { (\textbf{H}$_{ \theenumi }$) }
	\item	Alice is an information broker that intends to transmit some secret information to her network of $n$ agents, collectively named as Bob$_{ 0 }$, \dots, Bob$_{ n - 1 }$, who are all located in different geographical locations.
	\item	To accomplish her task, Alice employs a $( k, n )$ threshold QSS scheme, where $k > \frac { n } { 2 }$. For this purpose, she picks one of the typical procedures described in the literature to break the secret information into $n$ pieces in a way that any $k$ of them suffice to reconstruct the whole information, but any $k - 1$ or fewer do not.
	\item	Alice suspects that there may be a few rogue agents among Bob$_{ 0 }$, \dots, Bob$_{ n - 1 }$. Nonetheless, Alice is confident that there are at least $k$ loyal agents. In view of the constraint $k > \frac { n } { 2 }$, this means that the majority of her agents are loyal.
	\item	If Alice, during the verification phase, finds out that even one piece of information has been corrupted, it aborts the protocol. Otherwise, the execution of the protocol proceeds accordingly.
	\item	Upon successful completion of the DPVQSS protocol, all loyal agents must have obtained the secret information. It is acceptable though to also have a rogue agent obtain the secret information. This notion of successful completion of the DPVQSS protocol is inspired from the analogous criterion employed in protocols designed for achieving Detectable Byzantine Agreement \cite{Andronikos2023a}, where all loyal generals must follow the commanding general’s order. In our setup, we require all loyal agents to recover Alice's secret information.
\end{enumerate}

A situation where the protocol has to be aborted may arise due to either noisy quantum channels or intentional disruption by a rogue agent or an external adversary such as Eve. Irrespective of the true cause, the most appropriate course of action is to stop the current execution and restart the entire process, after appropriate corrective measures are put in place.

\begin{definition} [Information Vectors] \label{def: Information Vectors}
	Alice divides her secret information into $n$ pieces.
\end{definition}
\begin{itemize}
	\item	
	Each such piece is represented by the \emph{partial information bit vector} $\mathbf{ s }_{ i }$, which is transmitted to Bob$_{ i }$, $0 \leq i \leq n - 1$:
	\begin{align}
		\label{eq: Partial Information Bit Vector s_i}
		\mathbf{ s }_{ i }
		=
		b_{ i, m - 1 } \dots b_{ i, 0 }
		\ ,
	\end{align}
	where $b_{ i, j } \in \mathbb{ B }$, $0 \leq i \leq n - 1$ and $0 \leq j \leq m - 1$.
	\item	
	Given the $n$ partial information bit vectors $\mathbf{ s }_{ 0 }, \dots, \mathbf{ s }_{ n - 1 }$, Alice constructs the \emph{aggregated information bit vector} $\mathbf{ s }$ as their concatenation in the following ordering:
	\begin{align}
		\label{eq: Aggregated Information Bit Vector s}
		\mathbf{ s }
		=
		\mathbf{ s }_{ n - 1 }
		\dots
		\mathbf{ s }_{ 0 }
		=
		\underbrace { \colorbox {WordAquaLighter40} { $b_{ n - 1, m - 1 } \dots b_{ n - 1, 0 }$ } }_{ \mathbf{ s }_{ n - 1 } }
		\cdots
		\underbrace { \colorbox {WordAquaLighter80} { $b_{ 0, m - 1 } \dots b_{ 0, 0 }$ } }_{ \mathbf{ s }_{ 0 } }
		\ .
	\end{align}
	\item	
	Without loss of generality, we assume that all partial information vectors are of equal length $m$, which implies that the aggregated information bit vector $\mathbf{ s }$ has length $n m$.
\end{itemize}

The above discussion show that the aggregated information bit vector can be conceptually partitioned into $n$ \emph{segments}, with segment $i$ containing the partial information vector that is intended for one for Bob$_{ i }$, $0 \leq i \leq n - 1$. The relative ordering of the $n$ segments is the one indicated by formula \eqref{eq: Aggregated Information Bit Vector s}. This partitioning induces a corresponding partition of the players' quantum input registers into segments.

\begin{definition} [Segments] \label{def: Segments}
	The notion of \emph{segments} is helpful for the efficient manipulation of bit vectors, so, here we introduce some related definitions.
\end{definition}
\begin{itemize}
	\item	
	Any bit vector $\mathbf{ b }$ containing $n m$ bits is thought of as consisting of $n$ \emph{segments}, $0 \leq i \leq n - 1$. The precise partitioning scheme is shown below:
	\begin{align}
		\label{eq: The i-th Segment of a Bit Vector}
		\mathbf { b }
		&=
		\underbrace { \colorbox {WordLightGreen} { $b_{ n - 1, m - 1 } \dots b_{ n - 1, 0 }$ } }_{ \text{ segment } n - 1 }
		\cdots
		\underbrace { \colorbox {WordLightGreen!25} { $b_{ 0, m - 1 } \dots b_{ 0, 0 }$ } }_{ \text{ segment } 0 }
		\ .
	\end{align}
	For succinctness, the $i^{ th }$ segment will also be written as $\mathbf{ b }_{ i }$. Consequently, $\mathbf{ b }$ can also be expressed as:
	\begin{align}
		\label{eq: Bit Vector as Concatenation of Segments}
		\mathbf { b }
		=
		\mathbf{ b }_{ n - 1 }
		\dots
		\mathbf{ b }_{ 0 }
		\ .
	\end{align}
	\item	
	In an entire analogous manner, we may view any quantum register containing $n m$ qubits as consisting of $n$ \emph{segments}, $0 \leq i \leq n - 1$:
	\begin{align}
		\label{eq: The i-th Segment of a Quantum Register}
		\ket { \mathbf { x } }
		&=
		\underbrace { \colorbox {RedPurple!48} { $\ket{ x_{ n - 1, m - 1 } } \dots \ket{ x_{ n - 1, 0 } }$ } }_{ \text{ segment } n - 1 }
		\cdots
		\underbrace { \colorbox {RedPurple!12} { $\ket{ x_{ 0, m - 1 } } \dots \ket{ x_{ 0, 0 } }$ } }_{ \text{ segment } 0 }
		\ .
	\end{align}
	\item	
	Given a partial information vector $\mathbf{ s }_{ i } = b_{ i, m - 1 } \dots b_{ i, 0 }$, we define the \emph{extended} partial information vector $\widetilde { \mathbf{ s }_{ i } }$ as follows:
	\begin{align}
		\label{eq: Extended Partial Information Bit Vector s_i}
		\widetilde { \mathbf{ s }_{ i } }
		=
		\underbrace { \colorbox {WordVeryLightTeal} { $0 \dots 0$ } }_{ \text{ segment } n - 1 }
		\cdots
		\underbrace { \colorbox {WordVeryLightTeal} { $0 \dots 0$ } }_{ \text{ segment } i + 1 }
		\underbrace { \colorbox {WordAquaLighter40} { $b_{ i, m - 1 } \dots b_{ i, 0 }$ } }_{ \text{ segment } i }
		\underbrace { \colorbox {WordVeryLightTeal} { $0 \dots 0$ } }_{ \text{ segment } i - 1 }
		\cdots
		\underbrace { \colorbox {WordVeryLightTeal} { $0 \dots 0$ } }_{ \text{ segment } 0 }
		\ .
	\end{align}
	It is clear from the above definition that, although the partial information vector $\mathbf{ s }_{ i }$ has length $m$, the extended partial information vector $\widetilde { \mathbf{ s }_{ i } }$ has length $n m$; its $i^{ th }$ segment is $\mathbf{ s }_{ i }$, and all other segments are filled with $0$.
	\item	
	Likewise, for a bit vector $\mathbf{ b }$ with $n m$ bits, we define its \emph{extended} $i^{ th }$ segment $\widetilde { \mathbf{ b }_{ i } }$ as:
	\begin{align}
		\label{eq: Extended i-th Segment of Information Bit Vector b}
		\widetilde { \mathbf{ b }_{ i } }
		=
		\underbrace { \colorbox {WordVeryLightTeal} { $0 \dots 0$ } }_{ \text{ segment } n - 1 }
		\cdots
		\underbrace { \colorbox {WordVeryLightTeal} { $0 \dots 0$ } }_{ \text{ segment } i + 1 }
		\underbrace { \colorbox {WordAquaLighter40} { $b_{ i, m - 1 } \dots b_{ i, 0 }$ } }_{ \text{ segment } i }
		\underbrace { \colorbox {WordVeryLightTeal} { $0 \dots 0$ } }_{ \text{ segment } i - 1 }
		\cdots
		\underbrace { \colorbox {WordVeryLightTeal} { $0 \dots 0$ } }_{ \text{ segment } 0 }
		\ .
	\end{align}
\end{itemize}

\subsection{The entanglement distribution scheme} \label{subsec: The Entanglement Distribution Scheme}

The physical realization of the DPVQSS protocol assumes a compound system made of $n + 1$ local quantum circuits. The protocol relies on the fact that the corresponding qubits of all quantum registers are linked in a maximal entangled state
depending on the phase of the protocol. Technically, this is achieved by following the Uniform Entanglement Distribution Scheme, explained in the next Definition \ref{def: Uniform Entanglement Distribution Scheme}.

\begin{definition} [Uniform Distribution Scheme] \label{def: Uniform Entanglement Distribution Scheme}
	The $r$-Uniform Distribution Scheme asserts that:
\end{definition}
\begin{itemize}
	\item	
	there are $r$ players and each player is endowed with a $p$-qubit register, and
	\item	
	the qubits in the $j^{ th }$ position, $0 \leq j \leq p - 1$, of these quantum registers are entangled in the $\ket{ GHZ_{ r } }$ state, or in the $\ket{ \Phi^{ + } }$ state when $r = 2$.
\end{itemize}

This scheme creates a powerful correlation among the registers because their corresponding qubits are maximally entangled in the $\ket{ GHZ_{ r } }$ state.

Any actual implementation of the DPVQSS protocol will require Alice, or another trusted party, to create decoys. The use of decoys has been extensively explored in the related literature and there are numerous article giving a clear exposition its use, such as \cite{Deng2008, Yang2009, Tseng2011, Chang2013, Hung2016, Ye2018, Wu2021, Hou2024}. Decoys aim to expose the presence of a possible eavesdropper. In most scenarios, the party that prepares the entangled tuples also produces decoys. These are randomly chosen from one of the $\ket{ 0 }, \ket{ 1 }, \ket{ + }, \ket{ - }$ states, and randomly inserted into the transmitted sequence(s). The scheme in Definition \ref{def: Uniform Entanglement Distribution Scheme} doesn't include decoys in an effort to facilitate the overall understanding of the protocol.

\section{Detailed analysis of the DPVQSS protocol} \label{sec: Detailed Analysis of the DPVQSS Protocol}

The execution of the DPVQSS protocol evolves in phases, which are thoroughly presented in the current section.

\subsection{Information distribution phase} \label{subsec: Information Distribution Phase}

The first phase is the information distribution phase, where Alice sends the partial information bit vector $\mathbf{ s }_{ i }$ to Bob$_{ i }$, $0 \leq i \leq n - 1$. Our protocol is able to achieve this information distribution task in just one step, by enabling Alice to transmit all partial information vectors to their intended recipients simultaneously in a completely parallel manner.

\begin{tcolorbox}
	[
	enhanced,
	breakable,
	grow to left by = 0.00 cm,
	grow to right by = 0.00 cm,
	colback = WordAquaLighter80,			
	enhanced jigsaw,						
	sharp corners,
	toprule = 0.01 pt,
	bottomrule = 0.01 pt,
	leftrule = 0.1 pt,
	rightrule = 0.1 pt,
	sharp corners,
	center title,
	fonttitle = \bfseries
	]
	\begin{center}
		{ \large \textbf{ The intuition behind the information distribution phase } }
	\end{center}
	
	Our protocol is entanglement-based, meaning that it relies on the purely quantum phenomenon of entanglement in order to encode secret information in the global state of the compound quantum system. The formal mathematical relation of the correlations among the constituent subsystems is given by expression \eqref{eq: Hadamard Entanglement Property}, which throughout this paper we refer to as the \textbf{Hadamard Entanglement Property}.
\end{tcolorbox}

The quantum circuit that implements the information distribution phase of our protocol, denoted by IDPQS is graphically depicted in Figure \ref{fig: The Quantum Circuit the Information Distribution Phase}.

\begin{tcolorbox}
	[
		enhanced,
		breakable,
		grow to left by = 0.00 cm,
		grow to right by = 0.00 cm,
		colback = WordVeryLightTeal!25,			
		enhanced jigsaw,						
		sharp corners,
		toprule = 1.0 pt,
		bottomrule = 1.0 pt,
		leftrule = 0.1 pt,
		rightrule = 0.1 pt,
		sharp corners,
		center title,
		fonttitle = \bfseries
	]
	\begin{figure}[H]
		\centering
		\begin{tikzpicture} [ scale = 0.800 ] 
			\begin{yquant}
				nobit AUX_B_0_0;
				nobit AUX_B_0_1;
				[ name = Bob$_{ 0 }$ ] qubits { $BIR_{ 0 }$ } BIR1;
				nobit AUX_B_0_2;
				[ name = space_0, register/minimum height = 8 mm ] nobit space_0;
				nobit AUX_B_n-1_0;
				nobit AUX_B_n-1_1;
				[ name = Bob$_{ n - 1 }$ ] qubits { $BIR_{ n - 1 }$ } BIRn;
				nobit AUX_B_n-1_2;
				[ name = space_n_2, register/minimum height = 8 mm ] nobit space_n_2;
				nobit AUX_A_0;
				nobit AUX_A_1;
				[ name = Alice ] qubits { $AIR$ } AIR;
				qubit { $AOR$: \ $\ket{ - }$ } AOR;
				nobit AUX_A_2;
				nobit AUX_A_3;
				[ name = Ph0, WordBlueDarker, line width = 0.250 mm, label = { [ label distance = 0.20 cm ] north: Initial State } ]
				barrier ( - ) ;
				[ draw = RedPurple, fill = RedPurple, radius = 0.7 cm ] box {\color{white} \Large \sf{U}$_{ \mathbf{ s } }$} (AIR - AOR);
				[ name = Ph1, WordBlueDarker, line width = 0.250 mm, label = { [ label distance = 0.20 cm ] north: State 1 } ]
				barrier ( - ) ;
				[ draw = WordBlueDarker, fill = WordBlueDarker, radius = 0.6 cm ] box {\color{white} \Large \sf{H}$^{ \otimes n m }$} BIR1;
				[ draw = WordBlueDarker, fill = WordBlueDarker, radius = 0.6 cm ] box {\color{white} \Large \sf{H}$^{ \otimes n m }$} BIRn;
				[ draw = WordBlueDarker, fill = WordBlueDarker, radius = 0.6 cm ] box {\color{white} \Large \sf{H}$^{ \otimes n m }$} AIR;
				[ name = Ph2, WordBlueDarker, line width = 0.250 mm, label = { [ label distance = 0.20 cm ] north: State 2 } ]
				barrier ( - ) ;
				[ line width = .250 mm, draw = white, fill = black, radius = 0.6 cm ] measure BIR1;
				[ line width = .250 mm, draw = white, fill = black, radius = 0.6 cm ] measure BIRn;
				[ line width = .350 mm, draw = white, fill = black, radius = 0.6 cm ] measure AIR;
				[ name = Ph3, WordBlueDarker, line width = 0.250 mm, label = { [ label distance = 0.20 cm ] north: Measurement } ]
				barrier ( - ) ;
				output { $\ket{ \mathbf{ b }_{ 0 } }$ } BIR1;
				output { $\ket{ \mathbf{ b }_{ n - 1 } }$ } BIRn;
				output { $\ket{ \mathbf{ a } }$ } AIR;
				\node [ below = 5.00 cm ] at (Ph0) { $\ket{ \psi_{ 0 } }$ };
				\node [ below = 5.00 cm ] at (Ph1) { $\ket{ \psi_{ 1 } }$ };
				\node [ below = 5.00 cm ] at (Ph2) { $\ket{ \psi_{ 2 } }$ };
				\node [ below = 5.00 cm ] at (Ph3) { $\ket{ \psi_{ f } }$ };
				\node
					[
						charlie,
						scale = 1.50,
						anchor = center,
						left = 0.70 cm of Bob$_{ 0 }$,
						label = { [ label distance = 0.00 cm ] north: Bob$_{ 0 }$ }
					]
					() { };
				\node
					[
						bob,
						scale = 1.50,
						anchor = center,
						left = 0.30 cm of Bob$_{ n - 1 }$,
						label = { [ label distance = 0.00 cm ] north: Bob$_{ n - 1 }$ }
					]
					() { };
				\node
					[
						alice,
						scale = 1.50,
						anchor = center,
						left = 0.70 cm of Alice,
						label = { [ label distance = 0.000 cm ] north: Alice }
					]
					() { };
				\begin{scope} [ on background layer ]
					\node [ above left = 1.250 cm and 0.425 cm of Bob$_{ n - 1 }$ ] { \LARGE \vdots };
					\node [ above right = - 0.30 cm and 0.60 cm of space_0, rectangle, fill = WordAquaLighter60, text width = 10.00 cm, align = center, minimum height = 10 mm ] { \bf Spatially Separated };
					\node [ above right = - 0.30 cm and 0.60 cm of space_n_2, rectangle, fill = WordAquaLighter60, text width = 10.00 cm, align = center, minimum height = 10 mm ] { \bf Spatially Separated };
				\end{scope}
			\end{yquant}
			\node
				[
					above right = 2.750 cm and 4.500 cm of Bob$_{ 0 }$,
					anchor = center,
					shade,
					top color = GreenTeal, bottom color = black,
					rectangle,
					text width = 11.00 cm,
					align = center
				]
				(Label)
				{ \color{white}
					Using the quantum circuit IDPQC, Alice distributes the partial information bit vectors to her networks of agents.
				};
			\node [ anchor = center, below = 1.00 cm of Alice ] (PhantomNode) { };
			\scoped [ on background layer ]
			\draw
				[ MagentaDark, -, >=stealth, line width = 0.75 mm, decoration = coil, decorate ]
				( $ (Alice.east) + ( 0.5 mm, 0 mm ) $ ) node [ circle, fill, minimum size = 1.5 mm ] () {} -- ( $ (Bob$_{ n - 1 }$.east) + ( 0.5 mm, 0 mm ) $ ) node [ circle, fill, minimum size = 1.5 mm ] () {} -- ( $ (Bob$_{ 0 }$.east) + ( 0.5 mm, 0 mm ) $ ) node [ circle, fill, minimum size = 1.5 mm ] () {};
		\end{tikzpicture}
		\caption{This figure shows the IDPQC circuit that embeds Alice's secret $\mathbf{ s }$ into the global state of the system. Although the private circuits operated upon by Alice and her agents are spatially separated, they are linked due to entanglement, as indicated by the wavy magenta line connecting their input registers. The states $\ket{ \psi_{ 0 } }$, $\ket{ \psi_{ 1 } }$, $\ket{ \psi_{ 2 } }$, and $\ket{ \psi_{ f } }$ describe the evolution of the compound system.}
		\label{fig: The Quantum Circuit the Information Distribution Phase}
	\end{figure}
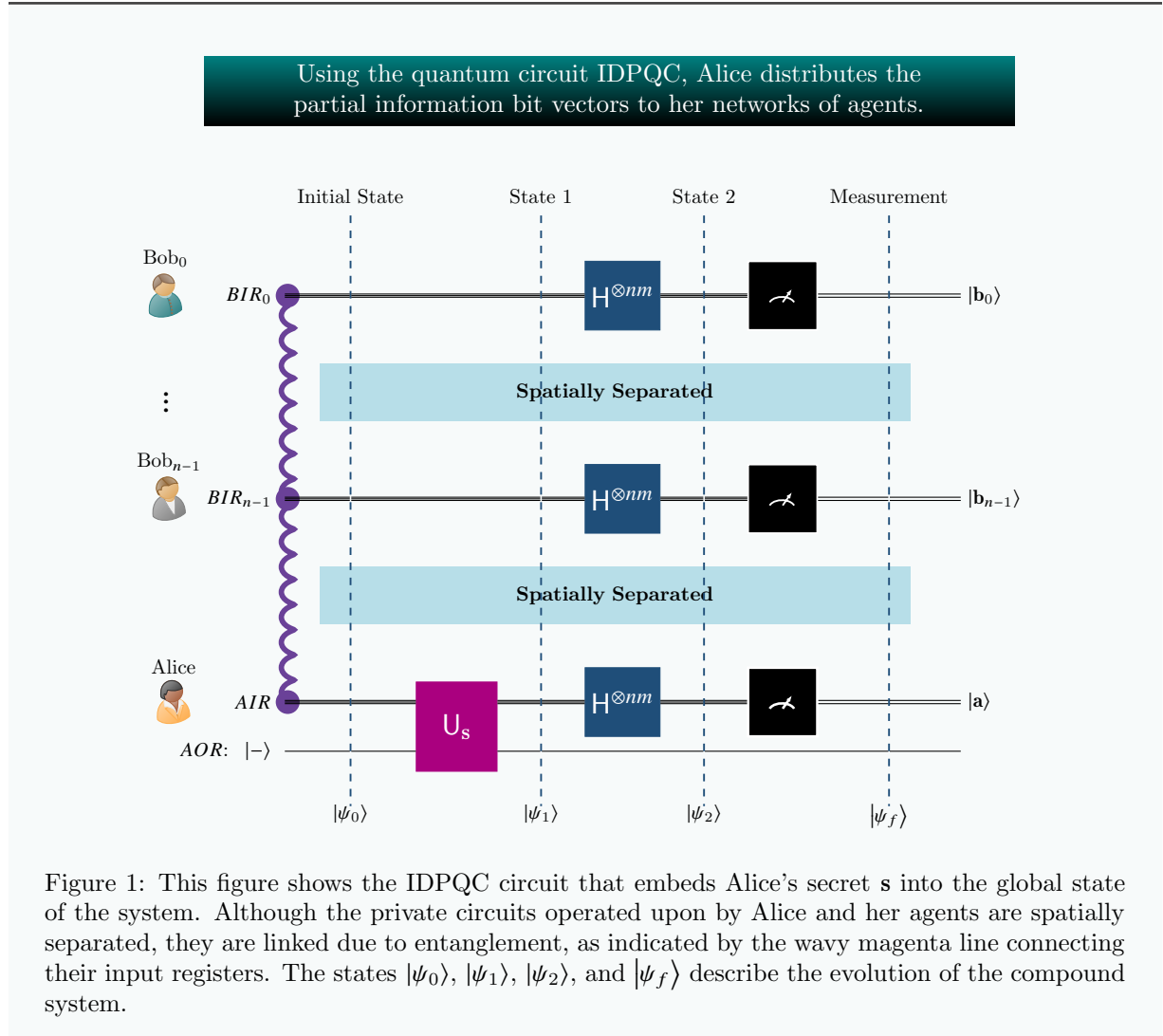
\end{tcolorbox}

In the quantum circuit of Figure \ref{fig: The Quantum Circuit the Information Distribution Phase}, the following conventions are observed.

\begin{itemize}
	\item	
	IDPQS, like all the quantum circuits demonstrated in this work, follows the Qiskit \cite{Qiskit2024} way of ordering qubits, i.e., the little-endian qubit indexing convention, where the least significant qubit is at the top of the figure and the most significant at the bottom.
	\item	
	$AIR$ is Alice's quantum input register that contains $n m$ qubits. $AOR$ is Alice's single-qubit output register that is initialized to state $\ket{ - }$.
	\item	
	$BIR_{ i }$ is the input register in Bob$_{ i }$'s local private quantum circuit, $0 \leq i \leq n - 1$, also having $n m$ qubits.
	\item	
	The $n + 1$ input registers $AIR$, $BIR_{ 0 }$, \dots, $BIR_{ n - 1 }$ adhere to the $r$-Uniform Distribution Scheme (recall Definition \ref{def: Uniform Entanglement Distribution Scheme}) with $r = n + 1$.
	\item	
	$U_{ \mathbf{ s } }$ is Alice's unitary transform. Its precise mathematical expression is given by equation \eqref{eq: Alice's Unitary Transform U_s}. Its purpose is to enable Alice to encode the aggregated information bit vector $\mathbf{ s }$ into the global state of the compound system.
	\item	
	$H^{ \otimes n m }$ is the $n m$-fold Hadamard transform, which is necessary for the decoding of the embedded information.
\end{itemize}

Invoking \eqref{eq: r-Fold Extended General GHZ_n State} with $p = n m$, we may express the initial state $\ket{ \psi_{ 0 } }$ of the IDPQC quantum circuit as shown below

\begin{align}
	\label{eq: Initial State of IDPQC}
	\ket{ \psi_{ 0 } }
	=
	2^{ - \frac { n m } { 2 } }
	\sum_{ \mathbf{ x } \in \mathbb{ B }^{ n m } }
	\
	\ket{ - }_{ A }
	\
	\ket{ \mathbf{ x } }_{ A }
	\
	\ket{ \mathbf{ x } }_{ n - 1 }
	\
	\dots
	\
	\ket{ \mathbf{ x } }_{ 0 }
	\ .
\end{align}

Note that in order to make clear which player we refer to, we employ the subscript $A$ for Alice, and the subscripts $0 \leq i \leq n - 1$, for Bob$_{ i }$.

The protocol proceeds by having Alice act on her private quantum circuits via their secret unitary transform $U_{ \mathbf{ s } }$. In that way, Alice embeds her secret information into the entangled state of the composite circuit. Alice's unitary transform $U_{ \mathbf{ s } }$ follows the typical scheme $U_{ \mathbf{ s } } \colon \ket{ y } \ \ket{ \mathbf{ x } }$ $\rightarrow$ $\ket{ y \oplus \left( \mathbf{ s } \bullet \mathbf{ x } \right) } \ \ket{ \mathbf{ x } }$, which can be explicitly written as

\begin{align}
	\label{eq: Alice's Unitary Transform U_s}
	U_{ \mathbf{ s } }
	&\colon
	\ket{ - }_{ A }
	\
	\ket{ \mathbf{ x } }_{ A }
	\rightarrow
	( - 1 )^{ \mathbf{ s } \bullet \mathbf{ x } }
	\
	\ket{ - }_{ A }
	\
	\ket{ \mathbf{ x } }_{ A }
	\ .
\end{align}

This causes the quantum circuit to enter the next state $\ket{ \psi_{ 1 } }$.

\begin{align}
	\label{eq: State 1 of IDPQC}
	\ket{ \psi_{ 1 } }
	&=
	2^{ - \frac { n m } { 2 } }
	\sum_{ \mathbf{ x } \in \mathbb{ B }^{ n m } }
	\
	\left(
	U_{ \mathbf{ s } }
	\
	\ket{ - }_{ A }
	\
	\ket{ \mathbf{ x } }_{ A }
	\right)
	\
	\ket{ \mathbf{ x } }_{ n - 1 }
	\
	\dots
	\
	\ket{ \mathbf{ x } }_{ 0 }
	\nonumber \\
	&\hspace{ - 0.100 cm } \overset { \eqref{eq: Alice's Unitary Transform U_s} } { = }
	2^{ - \frac { n m } { 2 } }
	\sum_{ \mathbf{ x } \in \mathbb{ B }^{ n m } }
	\
	( - 1 )^{ \mathbf{ s } \bullet \mathbf{ x } }
	\
	\ket{ - }_{ A }
	\
	\ket{ \mathbf{ x } }_{ A }
	\
	\ket{ \mathbf{ x } }_{ n - 1 }
	\
	\dots
	\
	\ket{ \mathbf{ x } }_{ 0 }
	\ .
\end{align}

Intuitively, equation \eqref{eq: State 1 of IDPQC} shows that Alice has encoded her secret aggregated bit vector $\mathbf{ s }$ into the relative phase of the compound system. Subsequently, all the players must collectively extract this information in a concerted way, specifically, by all applying the $n m$-fold Hadamard transform to their input registers, as shown in Figure \ref{fig: The Quantum Circuit the Information Distribution Phase}. This action drives the circuit into its next state $\ket{ \psi_{ 2 } }$.

\begin{align}
	\label{eq: State 2 of IDPQC - 1}
	\ket{ \psi_{ 2 } }
	&\hspace{-0.08 cm}\overset { \eqref{eq: State 1 of IDPQC} } { = }
	2^{ - \frac { n m } { 2 } }
	\sum_{ \mathbf{ x } \in \mathbb{ B }^{ n m } }
	\
	( - 1 )^{ \mathbf{ s } \bullet \mathbf{ x } }
	\
	\ket{ - }_{ A }
	\
	H^{ \otimes n m }
	\ket{ \mathbf{x} }_{ A }
	\
	H^{ \otimes n m }
	\ket{ \mathbf{ x } }_{ n - 1 }
	\
	\dots
	\
	H^{ \otimes n m }
	\ket{ \mathbf{ x } }_{ 0 }
	\ .
\end{align}

Setting $p = n m$ in equation \eqref{eq: Hadamard p-Fold Ket x}, we see that

\begin{align}
	\label{eq: Explicit Hadamard r-Fold Expansions - A}
	H^{ \otimes n m }
	\ket{ \mathbf{ x } }_{ A }
	&=
	2^{ - \frac { n m } { 2 } }
	\sum_{ \mathbf{ z } \in \mathbb{ B }^{ n m } }
	\
	( - 1 )^{ \mathbf{ z } \bullet \mathbf{ x } }
	\
	\ket{ \mathbf{ z } }_{ A }
	\ , \text{ and }
	\\
	\label{eq: Explicit Hadamard r-Fold Expansions - B}
	H^{ \otimes n m }
	\ket{ \mathbf{ x } }_{ i }
	&=
	2^{ - \frac { n m } { 2 } }
	\sum_{ \mathbf{ y }_{ i } \in \mathbb{ B }^{ n m } }
	\
	( - 1 )^{ \mathbf{ y }_{ i } \bullet \mathbf{ x } }
	\
	\ket{ \mathbf{ y }_{ i } }_{ i }
	\ , \ 0 \leq i \leq n - 1
	\ .
\end{align}

In view of the above relations, $\ket{ \psi_{ 2 } }$ can be rewritten as

\begin{align}
	\label{eq: State 2 of IDPQC - 2}
	\hspace{- 2.50 cm}
	\ket{ \psi_{ 2 } }
	=
	2^{ - \frac { n m ( n + 2 ) } { 2 } }
	\sum_{ \mathbf{ z } \in \mathbb{ B }^{ n m } }
	\sum_{ \mathbf{ y }_{ n - 1 } \in \mathbb{ B }^{ n m } }
	\dots
	\sum_{ \mathbf{ y }_{ 0 } \in \mathbb{ B }^{ n m } }
	\sum_{ \mathbf{ x } \in \mathbb{ B }^{ n m } }
	\
	( - 1 )^{ ( \mathbf{ s } \oplus \mathbf{ z } \oplus \mathbf{ y }_{ n - 1 } \oplus \dots \oplus \mathbf{ y }_{ 0 } )
		\bullet \mathbf{ x } }
	\
	\ket{ - }_{ A }
	\
	\ket{ \mathbf{ z } }_{ A }
	\
	\ket{ \mathbf{ y }_{ n - 1 } }_{ n - 1 }
	\
	\dots
	\
	\ket{ \mathbf{ y_{ 0 } } }_{ 0 }
	\ .
\end{align}

The characteristic inner product property for zero \eqref{eq: Inner Product Modulo $2$ Property For Zero} and nonzero \eqref{eq: Inner Product Modulo $2$ Property For NonZero} bit vectors enables us to simplify the previous formula because

\begin{itemize}
	\item	
	when $\mathbf{ s } \oplus \mathbf{ z } \oplus \mathbf{ y }_{ n - 1 } \oplus \dots \oplus \mathbf{ z }_{ 0 } \neq \mathbf{ 0 }$, or, equivalently, $\mathbf{ z } \oplus \mathbf{ y }_{ n - 1 } \oplus \dots \oplus \mathbf{ y }_{ 0 }$ $\neq$ $\mathbf{ s }$, the sum
	\begin{align*}
		\sum_{ \mathbf{ x } \in \mathbb{ B }^{ n m } }
		\
		( - 1 )^{ ( \mathbf{ s } \oplus \mathbf{ z } \oplus \mathbf{ y }_{ n - 1 } \oplus \dots \oplus \mathbf{ y }_{ 0 } ) \bullet \mathbf{ x } }
		\
		\ket{ - }_{ A }
		\
		\ket{ \mathbf{ z } }_{ A }
		\
		\ket{ \mathbf{ y }_{ n - 1 } }_{ n - 1 }
		\
		\dots
		\
		\ket{ \mathbf{ y_{ 0 } } }_{ 0 }
	\end{align*}
	appearing in \eqref{eq: State 2 of IDPQC - 2} becomes just $0$, and
	\item	
	when $\mathbf{ s } \oplus \mathbf{ z } \oplus \mathbf{ y }_{ n - 1 } \oplus \dots \oplus \mathbf{ z }_{ 0 } = \mathbf{ 0 }$, or, equivalently, $\mathbf{ z } \oplus \mathbf{ y }_{ n - 1 } \oplus \dots \oplus \mathbf{ y }_{ 0 }$ $=$ $\mathbf{ s }$, the sum
	\begin{align*}
		\sum_{ \mathbf{ x } \in \mathbb{ B }^{ n m } }
		\
		( - 1 )^{ ( \mathbf{ s } \oplus \mathbf{ z } \oplus \mathbf{ y }_{ n - 1 } \oplus \dots \oplus \mathbf{ y }_{ 0 } ) \bullet \mathbf{ x } }
		\
		\ket{ - }_{ A }
		\
		\ket{ \mathbf{ z } }_{ A }
		\
		\ket{ \mathbf{ y }_{ n - 1 } }_{ n - 1 }
		\
		\dots
		\
		\ket{ \mathbf{ y_{ 0 } } }_{ 0 }
	\end{align*}
	becomes
	\begin{align}
		\label{eq: Inner Product Modulo $2$ Property for IDPQC}
		2^{ n m }
		\
		\ket{ - }_{ A }
		\
		\ket{ \mathbf{ z } }_{ A }
		\
		\ket{ \mathbf{ y }_{ n - 1 } }_{ n - 1 }
		\
		\dots
		\
		\ket{ \mathbf{ y_{ 0 } } }_{ 0 }
		\ .
	\end{align}
\end{itemize}

This leads to the next much simpler form of $\ket{ \psi_{ 2 } }$.

\begin{align}
	\label{eq: State 2 of IDPQC - 3}
	\hspace{- 1.00 cm}
	\ket{ \psi_{ 2 } }
	=
	2^{ - \frac { n^{ 2 } m } { 2 } }
	\sum_{ \mathbf{ z } \in \mathbb{ B }^{ n m } }
	\sum_{ \mathbf{ y }_{ n - 1 } \in \mathbb{ B }^{ n m } }
	\dots
	\sum_{ \mathbf{ y }_{ 0 } \in \mathbb{ B }^{ n m } }
	\
	( - 1 )^{ ( \mathbf{ s } \oplus \mathbf{ z } \oplus \mathbf{ y }_{ n - 1 } \oplus \dots \oplus \mathbf{ y }_{ 0 } )
		\bullet \mathbf{ x } }
	\
	\ket{ - }_{ A }
	\
	\ket{ \mathbf{ z } }_{ A }
	\
	\ket{ \mathbf{ y }_{ n - 1 } }_{ n - 1 }
	\
	\dots
	\
	\ket{ \mathbf{ y_{ 0 } } }_{ 0 }
	\ ,
\end{align}

where

\begin{align}
	\label{eq: Hadamard Entanglement Property}
	\mathbf{ z }
	\oplus
	\mathbf{ y }_{ n - 1 }
	\oplus
	\dots
	\oplus
	\mathbf{ y }_{ 0 }
	=
	\mathbf{ s }
	\ .
\end{align}

The relation expressed by \eqref{eq: Hadamard Entanglement Property} is important because it quantifies how the states of the input registers of Alice and her agents are intertwined. This connection has arisen due to the initial entanglement of the input registers. Furthermore, this correlation encodes Alice's secret information. From now on, we shall refer to \eqref{eq: Hadamard Entanglement Property} as the \textbf{Hadamard Entanglement Property}.

The last step of the quantum part of the information distribution phase is the measurement of the input registers in the computational basis by all players. By doing so, the state of the compound system collapses to the final state $\ket{ \psi_{ f } }$.

\begin{align}
	\label{eq: Final State of IDPQC}
	\ket{ \psi_{ f } }
	&=
	\ket{ - }_{ A }
	\
	\ket{ \mathbf{ a } }_{ A }
	\
	\ket{ \mathbf{ b }_{ n - 1 } }_{ n - 1 }
	\
	\dots
	\
	\ket{ \mathbf{ b_{ 0 } } }_{ 0 }
	\ ,
	\quad
	\text{ where }
	\quad
	\mathbf{ a }
	\oplus
	\mathbf{ b }_{ n - 1 }
	\oplus
	\dots
	\oplus
	\mathbf{ b }_{ 0 }
	=
	\mathbf{ s }
	\ .
\end{align}

For the needs of our protocol, we cast relation \eqref{eq: Final State of IDPQC} in terms of segments. Denoting by $\mathbf{ a }_{ i }$, $\mathbf{ b }_{ 0, i }$, \dots , $\mathbf{ b }_{ n - 1, i }$, and $\mathbf{ s }_{ i }$ the $i^{ th }$ segments of $\mathbf{ a }$, $\mathbf{ b }_{ 0 }$, \dots , $\mathbf{ b }_{ n - 1 }$, and $\mathbf{ s }$, respectively, we write

\begin{align}
	\label{eq: Final Sum of IDPQC in Segment Form}
	&\hspace{0.525 cm}
	\mathbf{ a }_{ i }
	\oplus
	\mathbf{ b }_{ n - 1, i }
	\oplus
	\dots
	\oplus
	\mathbf{ b }_{ 0, i }
	=
	\mathbf{ s }_{ i }
	\ ,
	\text{ where }
	\ 0 \leq i \leq n - 1
	\ .
\end{align}

For the completion of the information distribution phase, our protocol will require communication among the players. This communication is entirely classical, taking place through pairwise authenticated classical channels.

\begin{definition} [Classical Communication Schemes] \label{def: Classical Communication Schemes}
	The classical communication schemes outlined next take care of the classical communication required within the framework of the DPVQSS protocol.
\end{definition}
\begin{itemize}
	\item	
	$C_{ A }^{ \downarrow }$ designates the transmission from each Bob$_{ i }$, $0 \leq i \leq n - 1$, to Alice of the measured contents $\mathbf{ b }_{ i }$ of his input register. This procedure is performed in a fully parallel manner, where all Bobs simultaneously send their data to Alice.
	\item	
	$C_{ i }^{ \downarrow }$ designates the transmission from Alice and every Bob$_{ j }$, $0 \leq j \neq i \leq n - 1$, to Bob$_{ i }$ of the $i^{ th }$ segments $\mathbf{ a }_{ i }$ and $\mathbf{ b }_{ j, i }$ of their input registers. These transmissions also occur simultaneously.
\end{itemize}

The following actions complete the information distribution phase.

\begin{enumerate}
	[ left = 0.500 cm, labelsep = 0.500 cm, start = 1 ]
	\renewcommand \labelenumi { (\theenumi) }
	\item	The communication schemes $C_{ 0 }^{ \downarrow }$, \dots, $C_{ n - 1 }^{ \downarrow }$ take place in parallel.
	\item	After this, Bob$_{ i }$, $0 \leq i \leq n - 1$, possesses the $i^{ th }$ segments from all input registers. In view of \eqref{eq: Final Sum of IDPQC in Segment Form}, this implies that Bob$_{ i }$ has obtained all the necessary information to recover $\mathbf{ s }_{ i }$.
\end{enumerate}

Thus, upon the successful completion of the information distribution phase, Alice has transmitted all the partial information vectors $\mathbf{ s }_{ i }$ to their intended recipients. At this point the situation is as described below.

\begin{enumerate}
	[ left = 0.500 cm, labelsep = 0.500 cm, start = 1 ]
	\renewcommand \labelenumi { (\theenumi) }
	\item	As intended by Alice in the first place, Bob$_{ i }$ knows the authorized information $\mathbf{ s }_{ i }$.
	\item	As is the norm in any QSS scheme, at this stage Bob$_{ i }$ has no knowledge of any other partial information vector $\mathbf{ s }_{ j }$, with $i \neq j$.
	\item	The entire setup is fully distributed, since all players are at different locations. Furthermore, all actions during the information distribution phase have transpired in parallel.
\end{enumerate}

Figure \ref{fig: Input Registers Upon Completion of the Information Distribution Phase} is meant to convey this situation. The arrows indicate that Alice has communicated the partial information vector $\mathbf{ s }_{ i }$, occupying the $i^{ th }$ segment of her secret vector $\mathbf{ s }$ to Bob$_{ i }$, $0 \leq i \leq n - 1$.

\begin{tcolorbox}
	[
		enhanced,
		breakable,
		grow to left by = 0.00 cm,
		grow to right by = 0.00 cm,
		colback = WordVeryLightTeal!25,			
		enhanced jigsaw,						
		sharp corners,
		toprule = 1.0 pt,
		bottomrule = 1.0 pt,
		leftrule = 0.1 pt,
		rightrule = 0.1 pt,
		sharp corners,
		center title,
		fonttitle = \bfseries
	]
	\begin{figure}[H]
		\centering
		\begin{tikzpicture} [ scale = 0.800, transform shape ]
			\node
				[
					alice,
					scale = 1.50,
					anchor = center,
					label = { [ label distance = 0.00 cm ] south: Alice }
				]
				(Alice) { };
			\matrix
				[
					right = 1.000 cm of Alice,
					matrix of nodes,
					column sep = 0.000 mm,
					nodes
					=
					{
						draw,
						line width = 0.100 mm,
						rectangle,
						anchor = center,			
						shade,
						outer color = RedPurple!75,
						inner color = white,
						minimum size = 12.00 mm,
						font = \scriptsize
					}
				]
				{
					\node { $b_{ n - 1, m - 1 }$ }; &
					\node { \dots }; &
					\node { $b_{ n - 1, 0 }$ };
					\\
				};
			\node [ right = 5.800 cm of Alice ] { \Large $\cdots$ };
			\matrix
				[
					right = 7.000 of Alice,
					matrix of nodes,
					column sep = 0.000 mm,
					nodes
					=
					{
						draw,
						line width = 0.100 mm,
						rectangle,
						anchor = center,			
						shade,
						outer color = GreenLighter2!75,
						inner color = white,
						minimum size = 12 mm,
						font = \scriptsize
					}
				]
				{
					\node { $b_{ 0, m - 1 }$ }; &
					\node { \dots }; &
					\node { $b_{ 0, 0 }$ };
					\\
				};
			\node
				[
					above = 3.000 cm of Alice,
					bob,
					scale = 1.50,
					anchor = center,
					label = { [ label distance = 0.00 cm ] south: Bob$_{ n - 1 }$ }
				]
				(Bob$_{ n - 1 }$) { };
			\matrix
				[
					right = 1.000 cm of Bob$_{ n - 1 }$,
					matrix of nodes,
					column sep = 0.000 mm,
					nodes
					=
					{
						draw,
						line width = 0.100 mm,
						rectangle,
						anchor = center,			
						shade,
						outer color = RedPurple!75,
						inner color = white,
						minimum size = 12 mm,
						font = \scriptsize
					}
				]
				{
					\node { $b_{ n - 1, m - 1 }$ }; &
					\node { \dots }; &
					\node { $b_{ n - 1, 0 }$ };
					\\
				};
			\node [ right = 5.800 cm of Bob$_{ n - 1 }$ ] { \Large $\cdots$ };
			\matrix
				[
					right = 7.000 of Bob$_{ n - 1 }$,
					matrix of nodes,
					column sep = 0.000 mm,
					nodes
					=
					{
						draw,
						line width = 0.100 mm,
						rectangle,
						anchor = center,			
						fill = WordVeryLightTeal,
						minimum size = 12 mm,
						font = \scriptsize
					}
				]
				{
					\node { $?$ }; &
					\node { \dots }; &
					\node { $?$ };
					\\
				};
			\node [ above = 1.250 cm of Bob$_{ n - 1 }$ ] (VDots) { \Large $\vdots$ };
			\node
				[
					above = 3.750 cm of Bob$_{ n - 1 }$,
					charlie,
					scale = 1.50,
					anchor = center,
					label = { [ label distance = 0.00 cm ] south: Bob$_{ 0 }$ }
				]
				(Bob$_{ 0 }$) { };
			\matrix
				[
					right = 1.000 cm of Bob$_{ 0 }$,
					matrix of nodes,
					column sep = 0.000 mm,
					nodes
					=
					{
						draw,
						line width = 0.100 mm,
						rectangle,
						anchor = center,			
						fill = WordVeryLightTeal,
						minimum size = 12 mm,
						font = \footnotesize
					}
				]
				{
					\node { $?$ }; &
					\node { \dots }; &
					\node { $?$ };
					\\
				};
			\node [ right = 5.800 cm of Bob$_{ 0 }$ ] { \Large $\cdots$ };
			\matrix
				[
					right = 7.000 of Bob$_{ 0 }$,
					matrix of nodes,
					column sep = 0.000 mm,
					nodes
					=
					{
						draw,
						line width = 0.100 mm,
						rectangle,
						anchor = center,			
						shade,
						outer color = GreenLighter2!75,
						inner color = white,
						minimum size = 12 mm,
						font = \footnotesize
					}
				]
				{
					\node { $b_{ 0, m - 1 }$ }; &
					\node { \dots }; &
					\node { $b_{ 0, 0 }$ };
					\\
				};
			\node
				[
					above right = 2.000 cm and 6.250 cm of Bob$_{ 0 }$,
					anchor = center,
					shade,
					top color = GreenTeal, bottom color = black,
					rectangle,
					text width = 9.00 cm,
					align = center
				]
				(Label)
				{ \color{white}
					\large The situation upon completion of the information distribution phase
				};
			\begin{scope} [ on background layer ]
				\draw
				[ ->, WordBlueLight,>=stealth, line width = 3.0 mm ]
				( $ (Alice.east) + ( 3.450 cm, 0 ) $ ) -- ( $ (Bob$_{ n - 1 }$.east) + ( 3.450 cm, -0.750 cm) $ );
				\draw
				[ ->, WordBlueLight,>=stealth, line width = 3.0 mm ]
				( $ (Alice.east) + ( 9.400 cm, 0 ) $ ) -- ( $ (Bob$_{ 0 }$.east) + ( 9.400 cm, -0.750 cm) $ );
			\end{scope}
			\node [ anchor = west, below = 0.500 cm of Alice ] (PhantomNode1) { };
		\end{tikzpicture}
		\caption{This figure demonstrates the situation upon successful completion of the information distribution phase. Alice has transmitted the partial information vector $\mathbf{ s }_{ i }$ to the intended recipient Bob$_{ i }$, $0 \leq i \leq n - 1$. At this point Bob$_{ i }$ only knows the authorized information $\mathbf{ s }_{ i }$ and is oblivious of any other information vector. The symbol "?" is used to convey the fact that the contents of all other segments in Bob$_{ i }$'s register are random.}
		\label{fig: Input Registers Upon Completion of the Information Distribution Phase}
	\end{figure}
\end{tcolorbox}

\subsection{Verification of the distribution phase} \label{subsec: Verification of the Distribution Phase}

Cryptographic protocols must consider not only the notorious eavesdropper Eve, but also the possible existence of rogue agents. As we emphasized in Section \ref{sec: An Overview of the DPVQSS Protocol}, the DPVQSS protocol implements a $( k, n )$ threshold QSS scheme, where Alice is certain that there are at least $k$ loyal agents, but suspects that among Bob$_{ 0 }$, \dots, Bob$_{ n - 1 }$ there may be a few rogue agents. Alice must ensure that upon successful completion of the DPVQSS protocol, all loyal agents will have obtained the secret information, while, at the same time, acknowledging the possibility that a disloyal agent might also obtain the secret information. To make sure that all loyal agents obtain the secret information in the presence of clandestine traitors, Alice must verify that all the partial information vectors have accurately reached their intended recipients.

Before we explain how the verification is performed, it is expedient to clarify a subtle point. Admittedly, it is possible for any rogue agent to sabotage the protocol during the information distribution phase. Even a single disloyal Bob, say Bob$_{ l }$, can prevent all other Bobs from obtaining the correct information; all he has to do is to send false or random data during the $C_{ 0 }^{ \downarrow }$, \dots, $C_{ n - 1 }^{ \downarrow }$ communication schemes. However, assuming that Bob$_{ l }$ is rational, which is the de facto assumption in Game Theory, this course of action entails two disadvantages.

\begin{enumerate}
	[ left = 0.500 cm, labelsep = 0.500 cm, start = 1 ]
	\renewcommand \labelenumi { (\theenumi) }
	\item	Alice will infer the presence of at least one traitor and she will abort the protocol.
	\item	Disloyal Bob$_{ l }$ will only obtain the partial information bit vector $\mathbf{ s }_{ l }$ and miss the opportunity to get additional information, perhaps even the whole secret.
\end{enumerate}

In contrast, by accurately reporting the measured contents of his register and enabling Alice to successfully verify the initial distribution of partial secrets, Bob$_{ l }$ stands a chance to obtain the complete information. Ergo, this has to be the chosen action of a rational Bob$_{ l }$.

\begin{tcolorbox}
	[
	enhanced,
	breakable,
	grow to left by = 0.00 cm,
	grow to right by = 0.00 cm,
	colback = WordAquaLighter80,			
	enhanced jigsaw,						
	sharp corners,
	toprule = 0.01 pt,
	bottomrule = 0.01 pt,
	leftrule = 0.1 pt,
	rightrule = 0.1 pt,
	sharp corners,
	center title,
	fonttitle = \bfseries
	]
	\begin{center}
		{ \large \textbf{ The intuition behind the verification } }
	\end{center}
	
	The DPVQSS protocol must ensure that every loyal Bob obtains the secret. If all the pieces of information sent by Alice during the previous phase have accurately reached their intended recipients, this will certainly be achieved. Thus, Alice must verify that this is indeed the case. To do that she asks every Bob$_{ i }$, $0 \leq i \leq n - 1$, to send her back the partial information bit vector $\mathbf{ s }_{ i }$ he has received in the previous phase. So, in this phase, the information flows in the opposite direction. At the end of this phase, if Alice recovers the aggregated information bit vector $\mathbf{ s }$, she authorizes the continuation of the protocol. If not, she announces that the protocol must be aborted. We emphasize that in such an eventuality, even if all traitors collaborate, they still cannot reconstruct the secret because $k > \frac { n } { 2 }$.
\end{tcolorbox}

Technically, this is implemented by having each Bob$_{ i }$ embed the extended partial information vector $\widetilde { \mathbf{ s }_{ i } }$, $0 \leq i \leq n - 1$, into the global state of the compound system. Formula \eqref{eq: Extended Partial Information Bit Vector s_i} implies that

\begin{align} \label{eq: Aggregated Information Bit Vector as the XOR of Extended s_i}
	\mathbf{ s }
	=
	\widetilde { \mathbf{ s }_{ n - 1 } }
	\oplus
	\dots
	\oplus
	\widetilde { \mathbf{ s }_{ 0 } }
	\ .
\end{align}

The quantum circuit responsible for the verification of the preceding distribution of partial secrets, denoted by VDPQS, is shown in Figure \ref{fig: The Quantum Circuit the Verification of the Distribution Phase}. Below, we explain the main differences between the circuit in Figure \ref{fig: The Quantum Circuit the Verification of the Distribution Phase} and the one in Figure \ref{fig: The Quantum Circuit the Information Distribution Phase}.

\begin{itemize}
	\item	
	Now Bob$_{ i }$'s, $0 \leq i \leq n - 1$, private circuit also contains the output register $BOR_{ i }$ and the unitary transform $U_{ \widetilde { \mathbf{ s }_{ i } } }$. $BOR_{ i }$ is initialized to $\ket{ - }$ and $U_{ \widetilde { \mathbf{ s }_{ i } } }$, given by \eqref{eq: Bob_i's Unitary Transform U_s_i}, embeds $\widetilde { \mathbf{ s }_{ i } }$ into the global state of the entangled system.
	\item	
	The output register $AOR$ and the unitary transform $U_{ \mathbf{ s } }$ are now absent from Alice's circuit.
	\item	
	These differences signify that the information flow is now from the agents to Alice.
\end{itemize}

The initial state $\ket{ \psi_{ 0 } }$ of the VDPQC quantum circuit is

\begin{align}
	\label{eq: Initial State of VDPQC}
	\ket{ \psi_{ 0 } }
	=
	2^{ - \frac { n m } { 2 } }
	\sum_{ \mathbf{ x } \in \mathbb{ B }^{ n m } }
	\
	\ket{ \mathbf{ x } }_{ A }
	\
	\ket{ - }_{ n - 1 }
	\
	\ket{ \mathbf{ x } }_{ n - 1 }
	\
	\dots
	\
	\ket{ - }_{ 0 }
	\
	\ket{ \mathbf{ x } }_{ 0 }
	\ .
\end{align}

Bob$_{ i }$'s, $0 \leq i \leq n - 1$, unitary transform $U_{ \widetilde { \mathbf{ s }_{ i } } }$ implements the standard scheme $U_{ \widetilde { \mathbf{ s }_{ i } } } \colon \ket{ y } \ \ket{ \mathbf{ x } }$ $\rightarrow$ $\ket{ y \oplus \left( \widetilde { \mathbf{ s }_{ i } } \bullet \mathbf{ x } \right) } \ \ket{ \mathbf{ x } }$ and can be more conveniently as

\begin{align}
	\label{eq: Bob_i's Unitary Transform U_s_i}
	U_{ \widetilde { \mathbf{ s }_{ i } } }
	&\colon
	\ket{ - }_{ i }
	\
	\ket{ \mathbf{ x } }_{ i }
	\rightarrow
	( - 1 )^{ \widetilde { \mathbf{ s }_{ i } } \bullet \mathbf{ x } }
	\
	\ket{ - }_{ i }
	\
	\ket{ \mathbf{ x } }_{ i }
	\ .
\end{align}

The combined actions of Bob$_{ 0 }$, \dots, Bob$_{ n - 1 }$ lead to the state $\ket{ \psi_{ 1 } }$.

\begin{align}
	\label{eq: State 1 of VDPQC}
	\ket{ \psi_{ 1 } }
	&=
	2^{ - \frac { n m } { 2 } }
	\sum_{ \mathbf{ x } \in \mathbb{ B }^{ n m } }
	\
	\ket{ \mathbf{ x } }_{ A }
	\
	\left(
	U_{ \widetilde { \mathbf{ s }_{ n - 1 } } }
	\
	\ket{ - }_{ n - 1 }
	\
	\ket{ \mathbf{ x } }_{ n - 1 }
	\right)
	\
	\dots
	\
	\left(
	U_{ \widetilde { \mathbf{ s }_{ 0 } } }
	\
	\ket{ - }_{ 0 }
	\
	\ket{ \mathbf{ x } }_{ 0 }
	\right)
	\nonumber \\
	&\hspace{ - 0.350 cm } \overset { \eqref{eq: Aggregated Information Bit Vector as the XOR of Extended s_i}, \eqref{eq: Bob_i's Unitary Transform U_s_i} } { = }
	2^{ - \frac { n m } { 2 } }
	\sum_{ \mathbf{ x } \in \mathbb{ B }^{ n m } }
	\
	( - 1 )^{ \mathbf{ s } \bullet \mathbf{ x } }
	\
	\ket{ \mathbf{ x } }_{ A }
	\
	\ket{ - }_{ n - 1 }
	\
	\ket{ \mathbf{ x } }_{ n - 1 }
	\
	\dots
	\
	\ket{ - }_{ 0 }
	\
	\ket{ \mathbf{ x } }_{ 0 }
	\ .
\end{align}

\begin{tcolorbox}
	[
		enhanced,
		breakable,
		grow to left by = 0.00 cm,
		grow to right by = 0.00 cm,
		colback = WordVeryLightTeal!25,			
		enhanced jigsaw,						
		sharp corners,
		toprule = 1.0 pt,
		bottomrule = 1.0 pt,
		leftrule = 0.1 pt,
		rightrule = 0.1 pt,
		sharp corners,
		center title,
		fonttitle = \bfseries
	]
	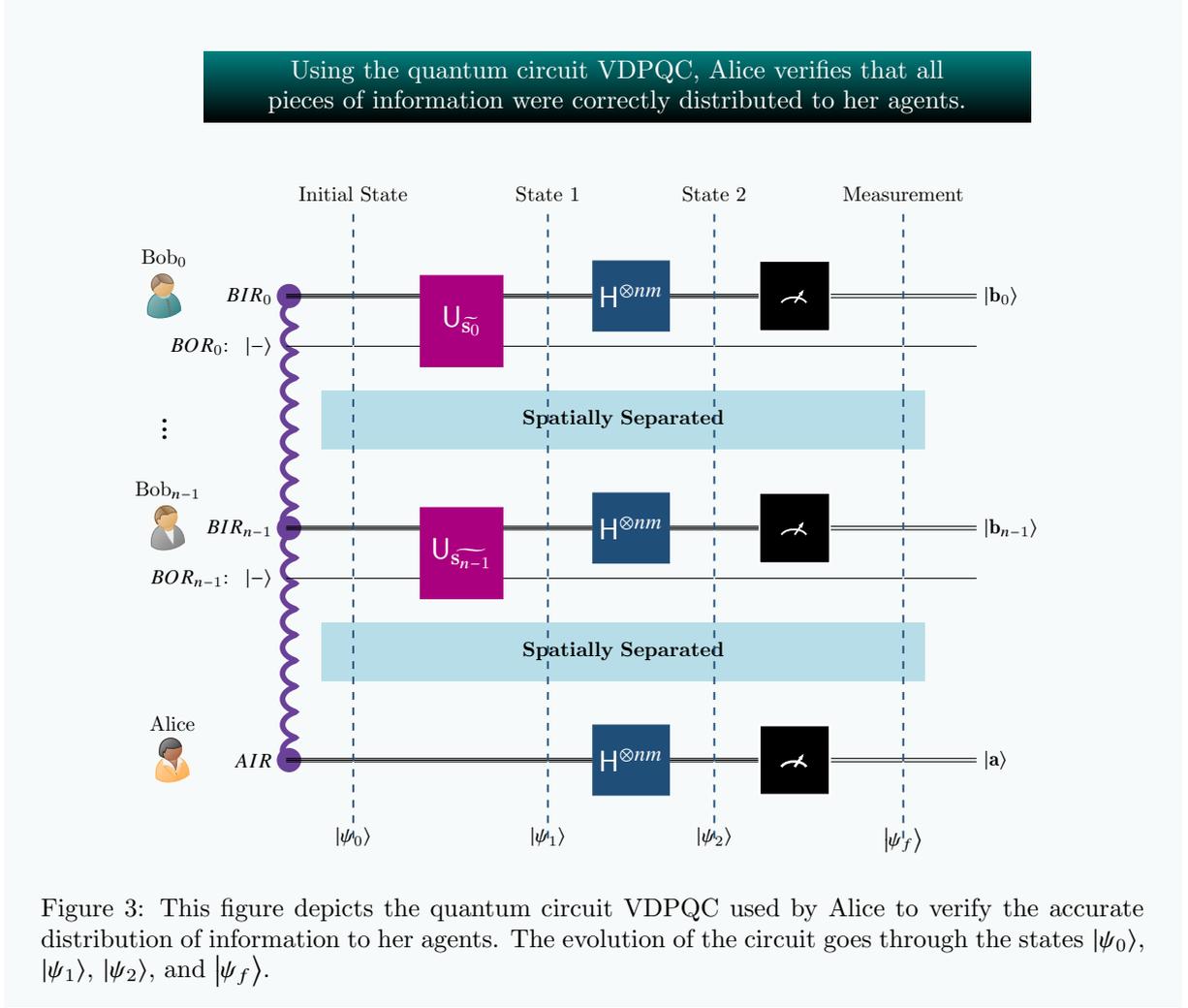
\begin{figure}[H]
		\centering
		\begin{tikzpicture} [ scale = 0.800 ] 
			\begin{yquant}
				nobit AUX_B_0_0;
				nobit AUX_B_0_1;
				[ name = Bob$_{ 0 }$ ] qubits { $BIR_{ 0 }$ } BIR1;
				qubit { $BOR_{ 0 }$: \ $\ket{ - }$ } BOR1;
				nobit AUX_B_0_2;
				[ name = space_0, register/minimum height = 8 mm ] nobit space_0;
				nobit AUX_B_n-1_0;
				nobit AUX_B_n-1_1;
				[ name = Bob$_{ n - 1 }$ ] qubits { $BIR_{ n - 1 }$ } BIRn;
				qubit { $BOR_{ n - 1 }$: \ $\ket{ - }$ } BORn;
				nobit AUX_B_n-1_2;
				[ name = space_n_2, register/minimum height = 8 mm ] nobit space_n_2;
				nobit AUX_A_0;
				nobit AUX_A_1;
				[ name = Alice ] qubits { $AIR$ } AIR;
				nobit AUX_A_2;
				nobit AUX_A_3;
				[ name = Ph0, WordBlueDarker, line width = 0.250 mm, label = { [ label distance = 0.20 cm ] north: Initial State } ]
				barrier ( - ) ;
				[ draw = RedPurple, fill = RedPurple, radius = 0.7 cm ] box {\color{white} \Large \sf{U}$_{ \widetilde { \mathbf{ s }_{ 0 } } }$} (BIR1 - BOR1);
				[ draw = RedPurple, fill = RedPurple, radius = 0.7 cm ] box {\color{white} \Large \sf{U}$_{ \widetilde { \mathbf{ s }_{ n - 1 } } }$} (BIRn - BORn);
				[ name = Ph1, WordBlueDarker, line width = 0.250 mm, label = { [ label distance = 0.20 cm ] north: State 1 } ]
				barrier ( - ) ;
				[ draw = WordBlueDarker, fill = WordBlueDarker, radius = 0.6 cm ] box {\color{white} \Large \sf{H}$^{ \otimes n m }$} BIR1;
				[ draw = WordBlueDarker, fill = WordBlueDarker, radius = 0.6 cm ] box {\color{white} \Large \sf{H}$^{ \otimes n m }$} BIRn;
				[ draw = WordBlueDarker, fill = WordBlueDarker, radius = 0.6 cm ] box {\color{white} \Large \sf{H}$^{ \otimes n m }$} AIR;
				[ name = Ph2, WordBlueDarker, line width = 0.250 mm, label = { [ label distance = 0.20 cm ] north: State 2 } ]
				barrier ( - ) ;
				[ line width = .250 mm, draw = white, fill = black, radius = 0.6 cm ] measure BIR1;
				[ line width = .250 mm, draw = white, fill = black, radius = 0.6 cm ] measure BIRn;
				[ line width = .350 mm, draw = white, fill = black, radius = 0.6 cm ] measure AIR;
				[ name = Ph3, WordBlueDarker, line width = 0.250 mm, label = { [ label distance = 0.20 cm ] north: Measurement } ]
				barrier ( - ) ;
				output { $\ket{ \mathbf{ b }_{ 0 } }$ } BIR1;
				output { $\ket{ \mathbf{ b }_{ n - 1 } }$ } BIRn;
				output { $\ket{ \mathbf{ a } }$ } AIR;
				\node [ below = 5.00 cm ] at (Ph0) { $\ket{ \psi_{ 0 } }$ };
				\node [ below = 5.00 cm ] at (Ph1) { $\ket{ \psi_{ 1 } }$ };
				\node [ below = 5.00 cm ] at (Ph2) { $\ket{ \psi_{ 2 } }$ };
				\node [ below = 5.00 cm ] at (Ph3) { $\ket{ \psi_{ f } }$ };
				\node
					[
						charlie,
						scale = 1.50,
						anchor = center,
						left = 0.70 cm of Bob$_{ 0 }$,
						label = { [ label distance = 0.00 cm ] north: Bob$_{ 0 }$ }
					]
					() { };
				\node
					[
						bob,
						scale = 1.50,
						anchor = center,
						left = 0.30 cm of Bob$_{ n - 1 }$,
						label = { [ label distance = 0.00 cm ] north: Bob$_{ n - 1 }$ }
					]
					() { };
				\node
					[
						alice,
						scale = 1.50,
						anchor = center,
						left = 0.70 cm of Alice,
						label = { [ label distance = 0.000 cm ] north: Alice }
					]
					() { };
				\begin{scope} [ on background layer ]
					\node [ above left = 1.250 cm and 0.425 cm of Bob$_{ n - 1 }$ ] { \LARGE \vdots };
					\node [ above right = - 0.30 cm and 0.60 cm of space_0, rectangle, fill = WordAquaLighter60, text width = 10.00 cm, align = center, minimum height = 10 mm ] { \bf Spatially Separated };
					\node [ above right = - 0.30 cm and 0.60 cm of space_n_2, rectangle, fill = WordAquaLighter60, text width = 10.00 cm, align = center, minimum height = 10 mm ] { \bf Spatially Separated };
				\end{scope}
			\end{yquant}
			\node
				[
					above right = 2.750 cm and 4.500 cm of Bob$_{ 0 }$,
					anchor = center,
					shade,
					top color = GreenTeal, bottom color = black,
					rectangle,
					text width = 11.00 cm,
					align = center
				]
				(Label)
			{ \color{white}
				Using the quantum circuit VDPQC, Alice verifies that all pieces of information were correctly distributed to her agents.
			};
			\node [ anchor = center, below = 1.00 cm of Alice ] (PhantomNode) { };
			\scoped [ on background layer ]
			\draw
				[ MagentaDark, -, >=stealth, line width = 0.75 mm, decoration = coil, decorate ]
				( $ (Alice.east) + ( 0.5 mm, 0 mm ) $ ) node [ circle, fill, minimum size = 1.5 mm ] () {} -- ( $ (Bob$_{ n - 1 }$.east) + ( 0.5 mm, 0 mm ) $ ) node [ circle, fill, minimum size = 1.5 mm ] () {} -- ( $ (Bob$_{ 0 }$.east) + ( 0.5 mm, 0 mm ) $ ) node [ circle, fill, minimum size = 1.5 mm ] () {};
		\end{tikzpicture}
		\caption{This figure depicts the quantum circuit VDPQC used by Alice to verify the accurate distribution of information to her agents. The evolution of the circuit goes through the states $\ket{ \psi_{ 0 } }$, $\ket{ \psi_{ 1 } }$, $\ket{ \psi_{ 2 } }$, and $\ket{ \psi_{ f } }$.}
		\label{fig: The Quantum Circuit the Verification of the Distribution Phase}
	\end{figure}
\end{tcolorbox}

Equation \eqref{eq: State 1 of VDPQC} implies that all agents have embedded their corresponding partial information vector into the composite system. In view of relation \eqref{eq: Aggregated Information Bit Vector as the XOR of Extended s_i}, this actually means that Alice's initial aggregated bit vector $\mathbf{ s }$ is encoded into the entangled circuit. To verify that this is indeed the case, all players must apply the $n m$-fold Hadamard transform to their input registers, as depicted in Figure \ref{fig: The Quantum Circuit the Verification of the Distribution Phase}. Accordingly, the system enters state $\ket{ \psi_{ 2 } }$.

\begin{align}
	\label{eq: State 2 of VDPQC - 1}
	\ket{ \psi_{ 2 } }
	&\hspace{-0.08 cm}\overset { \eqref{eq: State 1 of VDPQC} } { = }
	2^{ - \frac { n m } { 2 } }
	\sum_{ \mathbf{ x } \in \mathbb{ B }^{ n m } }
	\
	( - 1 )^{ \mathbf{ s } \bullet \mathbf{ x } }
	\
	H^{ \otimes n m }
	\ket{ \mathbf{x} }_{ A }
	\
	\ket{ - }_{ n - 1 }
	\
	H^{ \otimes n m }
	\ket{ \mathbf{ x } }_{ n - 1 }
	\
	\dots
	\
	\ket{ - }_{ 0 }
	\
	H^{ \otimes n m }
	\ket{ \mathbf{ x } }_{ 0 }
	\ .
\end{align}

Recalling the expansions given by equations \eqref{eq: Explicit Hadamard r-Fold Expansions - A} and \eqref{eq: Explicit Hadamard r-Fold Expansions - B}, we may write

\begin{align}
	\label{eq: State 2 of VDPQC - 2}
	\hspace{- 3.00 cm}
	\ket{ \psi_{ 2 } }
	=
	2^{ - \frac { n m ( n + 2 ) } { 2 } }
	\sum_{ \mathbf{ z } \in \mathbb{ B }^{ n m } }
	\sum_{ \mathbf{ y }_{ n - 1 } \in \mathbb{ B }^{ n m } }
	\dots
	\sum_{ \mathbf{ y }_{ 0 } \in \mathbb{ B }^{ n m } }
	\sum_{ \mathbf{ x } \in \mathbb{ B }^{ n m } }
	\
	( - 1 )^{ ( \mathbf{ s } \oplus \mathbf{ z } \oplus \mathbf{ y }_{ n - 1 } \oplus \dots \oplus \mathbf{ y }_{ 0 } )
		\bullet \mathbf{ x } }
	\
	\ket{ \mathbf{ z } }_{ A }
	\
	\ket{ - }_{ n - 1 }
	\
	\ket{ \mathbf{ y }_{ n - 1 } }_{ n - 1 }
	\
	\dots
	\
	\ket{ - }_{ 0 }
	\
	\ket{ \mathbf{ y_{ 0 } } }_{ 0 }
	\ .
\end{align}

Invoking the characteristic inner product property as expressed by formulae \eqref{eq: Inner Product Modulo $2$ Property For Zero} and \eqref{eq: Inner Product Modulo $2$ Property For NonZero}, we derive the following simplification of the previous formula

\begin{align}
	\label{eq: State 2 of VDPQC - 3}
	\hspace{- 1.50 cm}
	\ket{ \psi_{ 2 } }
	=
	2^{ - \frac { n^{ 2 } m } { 2 } }
	\sum_{ \mathbf{ z } \in \mathbb{ B }^{ n m } }
	\sum_{ \mathbf{ y }_{ n - 1 } \in \mathbb{ B }^{ n m } }
	\dots
	\sum_{ \mathbf{ y }_{ 0 } \in \mathbb{ B }^{ n m } }
	\
	( - 1 )^{ ( \mathbf{ s } \oplus \mathbf{ z } \oplus \mathbf{ y }_{ n - 1 } \oplus \dots \oplus \mathbf{ y }_{ 0 } )
		\bullet \mathbf{ x } }
	\
	\ket{ \mathbf{ z } }_{ A }
	\
	\ket{ - }_{ n - 1 }
	\
	\ket{ \mathbf{ y }_{ n - 1 } }_{ n - 1 }
	\
	\dots
	\
	\ket{ - }_{ 0 }
	\
	\ket{ \mathbf{ y_{ 0 } } }_{ 0 }
	\ ,
\end{align}

where the states of the input registers are correlated due to the Hadamard Entanglement Property and, consequently, satisfy relation \eqref{eq: Hadamard Entanglement Property}:

\begin{align}
	\tag{\ref{eq: Hadamard Entanglement Property}}
	\mathbf{ z }
	\oplus
	\mathbf{ y }_{ n - 1 }
	\oplus
	\dots
	\oplus
	\mathbf{ y }_{ 0 }
	=
	\mathbf{ s }
	\ .
\end{align}

By measuring the input registers, we arrive at the final state $\ket{ \psi_{ f } }$ of the compound system, which, in terms of the input registers, is identical to the final state \eqref{eq: Final State of IDPQC} at the end of the previous phase.

\begin{align}
	\tag{\ref{eq: Final State of IDPQC}}
	\ket{ \psi_{ f } }
	&=
	\ket{ \mathbf{ a } }_{ A }
	\
	\ket{ - }_{ n - 1 }
	\
	\ket{ \mathbf{ b }_{ n - 1 } }_{ n - 1 }
	\
	\dots
	\
	\ket{ - }_{ 0 }
	\
	\ket{ \mathbf{ b_{ 0 } } }_{ 0 }
	\ ,
	\quad
	\text{ where }
	\quad
	\mathbf{ a }
	\oplus
	\mathbf{ b }_{ n - 1 }
	\oplus
	\dots
	\oplus
	\mathbf{ b }_{ 0 }
	=
	\mathbf{ s }
	\ .
\end{align}

The $n$ agents conclude the current phase by simultaneously communicating with Alice through pairwise authenticated classical channels. The steps for the completion of the verification phase are listed below.

\begin{enumerate}
	[ left = 0.500 cm, labelsep = 0.500 cm, start = 1 ]
	\renewcommand \labelenumi { (\theenumi) }
	\item	The communication scheme $C_{ A }^{ \downarrow }$, in which every Bob$_{ i }$, $0 \leq i \leq n - 1$, sends to Alice the measured contents $\mathbf{ b }_{ i }$ of his input register, takes place in parallel.
	\item	Alice, having obtained all $n$ bit vectors $\mathbf{ b }_{ i }$, proceeds to compute their bitwise modulo-$2$ sum with her own bit vector $\mathbf{ a }$. As dictated by equation \eqref{eq: Final Sum of IDPQC}, the sum $\mathbf{ a } \oplus \mathbf{ b }_{ n - 1 } \oplus \dots \oplus \mathbf{ b }_{ 0 }$ must be equal to $\mathbf{ s }$.
\end{enumerate}

Hence, depending on the outcome of the verification process, Alice can undertake two diametrically opposite decisions.

\begin{enumerate}
	[ left = 0.500 cm, labelsep = 0.500 cm, start = 1 ]
	\renewcommand \labelenumi { (\theenumi) }
	\item	If the sum $\mathbf{ a } \oplus \mathbf{ b }_{ n - 1 } \oplus \dots \oplus \mathbf{ b }_{ 0 }$ is equal to $\mathbf{ s }$, Alice surmises that all agents Bob$_{ 0 }$, \dots, Bob$_{ n - 1 }$ have accurately received the intended piece of information. The DPVQSS protocol can safely proceed to the next phase, guaranteeing that all loyal agent will ultimately obtain the secret information.
	\item	If the sum $\mathbf{ a } \oplus \mathbf{ b }_{ n - 1 } \oplus \dots \oplus \mathbf{ b }_{ 0 }$ is not equal to $\mathbf{ s }$, Alice infers the existence of an adversary actively sabotaging the protocol. The only safe course of action is to abort the protocol.
	\item	This phase also transpires in a fully distributed and parallel manner because all players reside at different locations and the communication scheme $C_{ A }^{ \downarrow }$ is executed in parallel.
\end{enumerate}

If Alice greenlights its continuation, the protocol enters its third and final phase, explained in the next subsection \ref{subsec: Information Consolidation Phase}.

\subsection{Information Consolidation Phase} \label{subsec: Information Consolidation Phase}

Having progressed thus far, the DPVQSS protocol reaches its culmination in this last information consolidation phase. Upon completion of this phase, all loyal agents, which by assumption are at least $k$, will have obtained at least $k$ pieces of credible information, and will be able to recover the secret information, as was intended by Alice in the first place.

\begin{tcolorbox}
	[
	enhanced,
	breakable,
	grow to left by = 0.00 cm,
	grow to right by = 0.00 cm,
	colback = WordAquaLighter80,			
	enhanced jigsaw,						
	sharp corners,
	toprule = 0.01 pt,
	bottomrule = 0.01 pt,
	leftrule = 0.1 pt,
	rightrule = 0.1 pt,
	sharp corners,
	center title,
	fonttitle = \bfseries
	]
	\begin{center}
		{ \large \textbf{ The intuition behind the information consolidation phase } }
	\end{center}
	
	By following the procedure outlined in this subsection, an honest Bob$_{ i }$ will truthfully and accurately relay his partial information vector $\mathbf{ s }_{ i }$, representing the piece of information entrusted to him by Alice, to every other agent, honest or not. The fact that there are at least $k$ honest agents, implies that at least $k$ out of the $n$ pieces of information will reach every loyal agent. Thus, every loyal agent will reconstruct the secret information  because DPVQSS is a $( k, n )$ threshold QSS scheme.
\end{tcolorbox}

Implementation-wise, this is achieved by having each pair of distinct agents Bob$_{ i }$ and Bob$_{ j }$, where $0 \leq i < j \leq n - 1$, utilize the quantum circuit ICPQC$_{ i, j }$ in order to exchange their $\mathbf{ s }_{ i }$ and $\mathbf{ s }_{ j }$ information vectors. The pictorial representation of the ICPQC$_{ i, j }$ circuit is given in Figure \ref{fig: The Quantum Circuit ICPQC$_{ i, j }$ for the Information Consolidation Phase}.

\begin{tcolorbox}
	[
		enhanced,
		breakable,
		grow to left by = 0.00 cm,
		grow to right by = 0.00 cm,
		colback = WordVeryLightTeal!25,			
		enhanced jigsaw,						
		sharp corners,
		toprule = 1.0 pt,
		bottomrule = 1.0 pt,
		leftrule = 0.1 pt,
		rightrule = 0.1 pt,
		sharp corners,
		center title,
		fonttitle = \bfseries
	]
	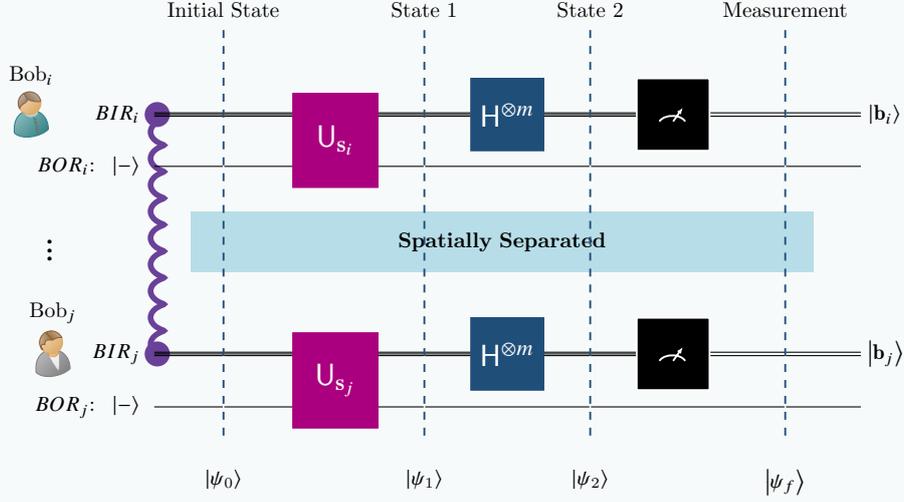
\begin{figure}[H]
		\centering
		\begin{tikzpicture} [ scale = 0.800 ] 
			\begin{yquant}
				nobit AUX_B_i_0;
				nobit AUX_B_i_1;
				[ name = Bob$_{ i }$ ] qubits { $BIR_{ i }$ } BIRi;
				qubit { $BOR_{ i }$: \ $\ket{ - }$ } BORi;
				nobit AUX_B_i_2;
				[ name = space_0, register/minimum height = 8 mm ] nobit space_0;
				nobit AUX_B_j_0;
				nobit AUX_B_j_1;
				[ name = Bob$_{ j }$ ] qubits { $BIR_{ j }$ } BIRj;
				qubit { $BOR_{ j }$: \ $\ket{ - }$ } BORj;
				nobit AUX_B_j_2;
				[ name = Ph0, WordBlueDarker, line width = 0.250 mm, label = { [ label distance = 0.20 cm ] north: Initial State } ]
				barrier ( - ) ;
				[ draw = RedPurple, fill = RedPurple, radius = 0.7 cm ] box {\color{white} \Large \sf{U}$_{ \mathbf{ s }_{ i } }$} (BIRi - BORi);
				[ draw = RedPurple, fill = RedPurple, radius = 0.7 cm ] box {\color{white} \Large \sf{U}$_{ \mathbf{ s }_{ j } }$} (BIRj - BORj);
				[ name = Ph1, WordBlueDarker, line width = 0.250 mm, label = { [ label distance = 0.20 cm ] north: State 1 } ]
				barrier ( - ) ;
				[ draw = WordBlueDarker, fill = WordBlueDarker, radius = 0.6 cm ] box {\color{white} \Large \sf{H}$^{ \otimes m }$} BIRi;
				[ draw = WordBlueDarker, fill = WordBlueDarker, radius = 0.6 cm ] box {\color{white} \Large \sf{H}$^{ \otimes m }$} BIRj;
				[ name = Ph2, WordBlueDarker, line width = 0.250 mm, label = { [ label distance = 0.20 cm ] north: State 2 } ]
				barrier ( - ) ;
				[ line width = .250 mm, draw = white, fill = black, radius = 0.6 cm ] measure BIRi;
				[ line width = .250 mm, draw = white, fill = black, radius = 0.6 cm ] measure BIRj;
				[ name = Ph3, WordBlueDarker, line width = 0.250 mm, label = { [ label distance = 0.20 cm ] north: Measurement } ]
				barrier ( - ) ;
				output { $\ket{ \mathbf{ b }_{ i } }$ } BIRi;
				output { $\ket{ \mathbf{ b }_{ j } }$ } BIRj;
				\node [ below = 3.750 cm ] at (Ph0) { $\ket{ \psi_{ 0 } }$ };
				\node [ below = 3.750 cm ] at (Ph1) { $\ket{ \psi_{ 1 } }$ };
				\node [ below = 3.750 cm ] at (Ph2) { $\ket{ \psi_{ 2 } }$ };
				\node [ below = 3.750 cm ] at (Ph3) { $\ket{ \psi_{ f } }$ };
				\node
					[
						charlie,
						scale = 1.50,
						anchor = center,
						left = 0.70 cm of Bob$_{ i }$,
						label = { [ label distance = 0.00 cm ] north: Bob$_{ i }$ }
					]
					() { };
				\node
					[
						bob,
						scale = 1.50,
						anchor = center,
						left = 0.30 cm of Bob$_{ j }$,
						label = { [ label distance = 0.00 cm ] north: Bob$_{ j }$ }
					]
					() { };
				\begin{scope} [ on background layer ]
					\node [ above left = 1.250 cm and 0.425 cm of Bob$_{ j }$ ] { \LARGE \vdots };
					\node [ above right = - 0.30 cm and 0.60 cm of space_0, rectangle, fill = WordAquaLighter60, text width = 10.00 cm, align = center, minimum height = 10 mm ] { \bf Spatially Separated };
				\end{scope}
			\end{yquant}
			\node
				[
					above right = 2.750 cm and 4.500 cm of Bob$_{ 0 }$,
					anchor = center,
					shade,
					top color = GreenTeal, bottom color = black,
					rectangle,
					text width = 11.00 cm,
					align = center
				]
				(Label)
			{ \color{white}
				Bob$_{ i }$ and Bob$_{ j }$ exchange their information vectors $\mathbf{ s }_{ i }$ and $\mathbf{ s }_{ j }$ with the use of the quantum circuit ICPQC$_{ i, j }$.
			};
			\node [ anchor = center, below = 1.00 cm of Bob$_{ j }$ ] (PhantomNode) { };
			\scoped [ on background layer ]
			\draw
				[ MagentaDark, -, >=stealth, line width = 0.75 mm, decoration = coil, decorate ]
				( $ (Bob$_{ j }$.east) + ( 0.5 mm, 0 mm ) $ ) node [ circle, fill, minimum size = 1.5 mm ] () {} -- ( $ (Bob$_{ i }$.east) + ( 0.5 mm, 0 mm ) $ ) node [ circle, fill, minimum size = 1.5 mm ] () {};
		\end{tikzpicture}
		\caption{For each pair of distinct agents Bob$_{ i }$ and Bob$_{ j }$, where $0 \leq i < j \leq n - 1$, the quantum circuit ICPQC$_{ i, j }$ realizes the exchange of their $\mathbf{ s }_{ i }$ and $\mathbf{ s }_{ j }$ information vectors. The evolution of the circuit goes through the states $\ket{ \psi_{ 0 } }$, $\ket{ \psi_{ 1 } }$, $\ket{ \psi_{ 2 } }$, and $\ket{ \psi_{ f } }$.}
		\label{fig: The Quantum Circuit ICPQC$_{ i, j }$ for the Information Consolidation Phase}
	\end{figure}
\end{tcolorbox}

The ICPQC$_{ i, j }$ circuit is considerably simpler than the IDPQC and VDPQC circuits; compare for instance Figure \ref{fig: The Quantum Circuit ICPQC$_{ i, j }$ for the Information Consolidation Phase} with Figures \ref{fig: The Quantum Circuit the Information Distribution Phase} and \ref{fig: The Quantum Circuit the Verification of the Distribution Phase}. The main differences of the former compared to the latter are summarized below.

\begin{itemize}
	\item	
	The ICPQC$_{ i, j }$ circuit is comprised of just two subcircuits, whereas both IDPQC and VDPQC circuits consist of $n + 1$ subcircuits.
	\item	
	Now Bob$_{ i }$ and Bob$_{ j }$'s output registers $BIR_{ i }$ and $BIR_{ j }$ contain $m$ qubits instead of $n m$, the reason being that the $\mathbf{ s }_{ i }$ and $\mathbf{ s }_{ j }$ information vectors have length $m$. These two input registers adhere to the $r$-Uniform Distribution Scheme of Definition \ref{def: Uniform Entanglement Distribution Scheme} with $r = 2$.
	\item	
	These differences illustrate that the information flow is now symmetrical and bidirectional, and both agents are on an equal footing.
	\item	
	The ICPQC$_{ i, j }$ circuits are independent of each other, and, thus, can all be executed in parallel.
\end{itemize}

The initial state $\ket{ \psi_{ 0 } }$ of the ICPQC$_{ i, j }$ circuit is

\begin{align}
	\label{eq: Initial State of ICPQC$_{ i, j }$}
	\ket{ \psi_{ 0 } }
	=
	2^{ - \frac { m } { 2 } }
	\sum_{ \mathbf{ x } \in \mathbb{ B }^{ n m } }
	\
	\ket{ - }_{ j }
	\
	\ket{ \mathbf{ x } }_{ j }
	\
	\ket{ - }_{ i }
	\
	\ket{ \mathbf{ x } }_{ i }
	\ .
\end{align}

The unitary transforms $U_{ \mathbf{ s }_{ i } }$ and $U_{ \mathbf{ s }_{ j } }$ employed by Bob$_{ i }$ and Bob$_{ j }$, $0 \leq i < j \leq n - 1$, are given by the following equations:

\begin{align}
	\label{eq: Bob_i,j's Unitary Transforms U_s_i,j}
	U_{ \mathbf{ s }_{ i } }
	\colon
	\ket{ - }_{ i }
	\
	\ket{ \mathbf{ x } }_{ i }
	\rightarrow
	( - 1 )^{ \mathbf{ s }_{ i } \bullet \mathbf{ x } }
	\
	\ket{ - }_{ i }
	\
	\ket{ \mathbf{ x } }_{ i }
	\quad
	\text{and}
	\quad
	U_{ \mathbf{ s }_{ j } }
	\colon
	\ket{ - }_{ j }
	\
	\ket{ \mathbf{ x } }_{ j }
	\rightarrow
	( - 1 )^{ \mathbf{ s }_{ j } \bullet \mathbf{ x } }
	\
	\ket{ - }_{ j }
	\
	\ket{ \mathbf{ x } }_{ j }
	\ .
\end{align}

Their actions lead to the state $\ket{ \psi_{ 1 } }$.

\begin{align}
	\label{eq: State 1 of ICPQC$_{ i, j }$}
	\hspace{-1.00 cm}
	\ket{ \psi_{ 1 } }
	=
	2^{ - \frac { m } { 2 } }
	\sum_{ \mathbf{ x } \in \mathbb{ B }^{ m } }
	\
	(
	U_{ \mathbf{ s }_{ j } }
	\
	\ket{ - }_{ j }
	\
	\ket{ \mathbf{ x } }_{ j }
	)
	\
	\left(
	U_{ \mathbf{ s }_{ i } }
	\
	\ket{ - }_{ i }
	\
	\ket{ \mathbf{ x } }_{ i }
	\right)
	\overset { \eqref{eq: Bob_i,j's Unitary Transforms U_s_i,j} } { = }
	2^{ - \frac { m } { 2 } }
	\sum_{ \mathbf{ x } \in \mathbb{ B }^{ m } }
	\
	( - 1 )^{ ( \mathbf{ s }_{ j } \oplus \mathbf{ s }_{ i } ) \bullet \mathbf{ x } }
	\
	\ket{ - }_{ j }
	\
	\ket{ \mathbf{ x } }_{ j }
	\
	\ket{ - }_{ i }
	\
	\ket{ \mathbf{ x } }_{ i }
	\ .
\end{align}

Interpreting the previous equation \eqref{eq: State 1 of ICPQC$_{ i, j }$}, we see that Bob$_{ i }$ and Bob$_{ j }$ have encoded the modulo-$2$ sum of their information vectors $\mathbf{ s }_{ i }$ and $\mathbf{ s }_{ j }$ into the relative phase of the compound system. To decode this information, both players must apply the $m$-fold Hadamard transform to their input registers,
which drives the system into state $\ket{ \psi_{ 2 } }$.

\begin{align}
	\label{eq: State 2 of ICPQC$_{ i, j }$ - 1}
	\ket{ \psi_{ 2 } }
	\overset { \eqref{eq: State 1 of ICPQC$_{ i, j }$} } { = }
	2^{ - \frac { m } { 2 } }
	\sum_{ \mathbf{ x } \in \mathbb{ B }^{ m } }
	\
	( - 1 )^{ ( \mathbf{ s }_{ j } \oplus \mathbf{ s }_{ i } ) \bullet \mathbf{ x } }
	\
	\ket{ - }_{ j }
	\
	H^{ \otimes m }
	\ket{ \mathbf{ x } }_{ j }
	\
	\ket{ - }_{ i }
	\
	H^{ \otimes m }
	\ket{ \mathbf{ x } }_{ i }
	\ ,
\end{align}

which, by equation \eqref{eq: Hadamard p-Fold Ket x} with $p = m$, becomes

\begin{align}
	\label{eq: State 2 of ICPQC$_{ i, j }$ - 2}
	\ket{ \psi_{ 2 } }
	=
	2^{ - \frac { 3 m } { 2 } }
	\sum_{ \mathbf{ y }_{ j } \in \mathbb{ B }^{ m } }
	\sum_{ \mathbf{ y }_{ i } \in \mathbb{ B }^{ m } }
	\sum_{ \mathbf{ x } \in \mathbb{ B }^{ m } }
	\
	( - 1 )^{ ( \mathbf{ s }_{ j } \oplus \mathbf{ s }_{ i } ) \bullet \mathbf{ x } }
	\
	\ket{ - }_{ j }
	\
	\ket{ \mathbf{ y }_{ j } }_{ j }
	\
	\ket{ - }_{ i }
	\
	\ket{ \mathbf{ y_{ i } } }_{ i }
	\ ,
\end{align}

which, due to \eqref{eq: Inner Product Modulo $2$ Property For Zero} and \eqref{eq: Inner Product Modulo $2$ Property For NonZero}, simplifies to

\begin{align}
	\label{eq: State 2 of ICPQC$_{ i, j }$ - 3}
	\ket{ \psi_{ 2 } }
	=
	2^{ - \frac { m } { 2 } }
	\sum_{ \mathbf{ y }_{ j } \in \mathbb{ B }^{ m } }
	\sum_{ \mathbf{ y }_{ i } \in \mathbb{ B }^{ m } }
	\
	( - 1 )^{ ( \mathbf{ s }_{ j } \oplus \mathbf{ s }_{ i } ) \bullet \mathbf{ x } }
	\
	\ket{ - }_{ j }
	\
	\ket{ \mathbf{ y }_{ j} }_{ j }
	\
	\ket{ - }_{ i }
	\
	\ket{ \mathbf{ y_{ i } } }_{ i }
	\ ,
\end{align}

where, as expected, the states of the input registers are intertwined due to entanglement, and satisfy the following special form of the Hadamard Entanglement Property:

\begin{align}
	\label{eq: Special Form of the Hadamard Entanglement Property}
	\mathbf{ y }_{ j }
	\oplus
	\mathbf{ y }_{ i }
	=
	\mathbf{ s }_{ j }
	\oplus
	\mathbf{ s }_{ i }
	\ .
\end{align}

By measuring their input registers, Bob$_{ i }$ and Bob$_{ j }$ arrive at the final state $\ket{ \psi_{ f } }$ of the compound system, which can also be perceived as a special case of \eqref{eq: Final State of IDPQC}.

\begin{align}
	\label{eq: Final State of ICPQC$_{ i, j }$}
	\ket{ \psi_{ f } }
	=
	\ket{ - }_{ j }
	\
	\ket{ \mathbf{ b }_{ j } }_{ j }
	\
	\ket{ - }_{ i }
	\
	\ket{ \mathbf{ b_{ i } } }_{ i }
	\ ,
	\quad
	\text{ where }
	\quad
	\mathbf{ b }_{ j }
	\oplus
	\mathbf{ b }_{ i }
	=
	\mathbf{ s }_{ j }
	\oplus
	\mathbf{ s }_{ i }
	\ .
\end{align}

This phase is concluded by having the agents undertake the following steps.

\begin{enumerate}
	[ left = 0.500 cm, labelsep = 0.500 cm, start = 1 ]
	\renewcommand \labelenumi { (\theenumi) }
	\item	Each pair of Bob$_{ i }$ and Bob$_{ j }$, $0 \leq i < j \leq n - 1$, exchange the measured contents of their input registers $\mathbf{ b }_{ i }$ and $\mathbf{ b }_{ j }$ through pairwise authenticated classical channels.
	\item	Following this communication exchange, Bob$_{ i }$ (Bob$_{ j }$) employs \eqref{eq: Final State of ICPQC$_{ i, j }$} to recover the secret information vector $\mathbf{ s }_{ j }$ ($\mathbf{ s }_{ i }$).
	\item	The whole process of communication exchange between pairs and secret information recovery takes place in parallel.
\end{enumerate}

The presence of rogue agents, cannot stop an honest Bob$_{ i }$ from truthfully and accurately transmitting his partial information vector $\mathbf{ s }_{ i }$, representing the piece of information entrusted to him by Alice, to every other agent. The fundamental assumption that there are at least $k$ loyal agents, means that at least $k$ out of the $n$ pieces of information will reach every loyal agent. Ergo, every loyal agent will reconstruct the secret information  because DPVQSS is a $( k, n )$ threshold QSS scheme.

\section{Efficiency and security analysis} \label{sec: Efficiency & Security Analysis}

The current section provides an analysis of the DPVQSS protocol in terms of efficiency and security.

\subsection{Efficiency} \label{subsec: Efficiency}

The most common metric for the qubit efficiency of quantum protocols is the ratio $\eta$ of the total number of transmitted ``useful'' classical bits to the total number of qubits employed by the protocol \cite{Tsai2011, Hwang2011}. To avoid any misconception, we point out that the standard approach in the literature is not to take into account the decoys in the calculation of $\eta$. In the case of the DPVQSS protocol that progresses in three phases, we shall compute the $\eta$ ratio for each phase.

\begin{enumerate}
	[ left = 0.500 cm, labelsep = 1.000 cm, start = 1 ]
	\renewcommand\labelenumi{(\textbf{E}$_\theenumi$)}
	\item	During the information distribution phase, Alice uses an input register containing $n m$ qubits and a single qubit output register, while Bob$_{ 0 }$, \dots, Bob$_{ n - 1 }$ each use an $n m$-qubit register, so that Alice can transmit $n m$ bits of information. Consequently, the $\eta_{ 1 }$ efficiency for this phase is
	\begin{align} \label{eq: IDP eta Efficiency}
		\eta_{ 1 }
		=
		\frac
		{ n m }
		{ ( n + 1 ) n m + 1 }
		\approx
		\frac
		{ n m }
		{ n^{ 2 } m }
		=
		\frac
		{ 1 }
		{ n }
		\ .
	\end{align}
	\item	During the verification phase, Alice uses an input register with $n m$ qubits, whereas each Bob$_{ i }$, $0 \leq i \leq n - 1$, utilizes an $n m$-qubit register and a single qubit output register. At the end of this phase, Alice obtains $n m$ bits of information at the end of this phase. Therefore, the $\eta_{ 2 }$ efficiency for this phase is
	\begin{align} \label{eq: VDP eta Efficiency}
		\eta_{ 2 }
		=
		\frac
		{ n m }
		{ ( n + 1 ) n m + n }
		\approx
		\frac
		{ n m }
		{ n^{ 2 } m }
		=
		\frac
		{ 1 }
		{ n }
		\ .
	\end{align}
	\item	In the last phase, every pair of distinct agents Bob$_{ i }$ and Bob$_{ j }$, $0 \leq i < j \leq n - 1$, use an $m$-qubit input register and a single qubit output register each, to exchange $m$ bits of information. Hence, the $\eta_{ 3 }$ efficiency for the information consolidation phase is
	\begin{align} \label{eq: ICP eta Efficiency}
		\eta_{ 3 }
		=
		\frac
		{ 2 m }
		{ 2 ( m + 1 ) }
		\approx
		\frac
		{ m }
		{ m }
		=
		100\%
		\ .
	\end{align}
\end{enumerate}

The above analysis demonstrates that the efficiency of the first two phases is not optimal, whereas the efficiency of the last phase is optimal. This antithesis can be attributed to two factors.

\begin{itemize}
	\item	
	The first two phases involve simultaneous communication among $n + 1$ players, in which all players participate equally and inherently in the transmission of information. In contrast, the third phase involves simultaneous communication among $2$ players.
	\item	
	In the first two phases the flow of information is unidirectional: from Alice to her agents in the first phase, and vice versa in the second phase. In the third phase the flow of information is bidirectional because both agents exchange information simultaneously.
\end{itemize}

\subsection{Security} \label{subsec: Security}

In this subsection we refer to Alice and her agents Bob$_{ 0 }$, \dots, Bob$_{ n - 1 }$ as the ``inside'' players, while all players are called ``external.'' Protocols like DPVQSS present an extra challenge when it comes to security analysis. This is because in addition the ever present Eve playing the role of external eavesdropper, there may well be inside traitors. In the literature there is a more or less standard approach to the security analysis (see \cite{Wolf2021, Renner2023} for more details), which we also follow here, making appropriate adjustments to account for our specific situation. Henceforth, we distinguish between \emph{external} and \emph{inside} attacks. By the former we understand attacks unleashed by Eve, and by the latter we mean attacks by one or more rogue Bobs.

There are some standard techniques that enhance the defensive capabilities of quantum protocols. One such technique, indispensable in our setup, is called the \emph{decoy technique} and is designed for \emph{eavesdropping detection}, that is, to expose the presence of a possible eavesdropper. According to this approach, the trusted party that produces and distributes the entangled tuples, also generates decoy qubits. These decoys are randomly chosen from one of the $\ket{ 0 }, \ket{ 1 }, \ket{ + }, \ket{ - }$ states, and are embedded into the transmission sequences in random positions. Eve, while tampering with the transmission sequences and being oblivious of the positions of the decoys, will introduce errors that the legitimate players can detect. This method has been thoroughly examined in the existing literature \cite{Deng2008, Yang2009, Tseng2011, Chang2013, Hung2016, Ye2018, Wu2021, Hou2024}.

Regarding entanglement-based protocols in particular, it is, perhaps, unnecessary to point out that their successful realization hinges upon the existence of entanglement. In the absence of entanglement, such protocol are doomed to utter failure. Thus, it is imperative to establish a reliable method for \emph{entanglement validation}. If entanglement is verified, the protocol continues to successfully carry out its designated objective. In the opposite case, the required entanglement is conspicuously absent. There are numerous causes for such an eventuality, such as noisy quantum channels or tampering from an adversary. In any event, the only sensible response is to abort the protocol and restart from scratch, after appropriate corrective measures are implemented. Entanglement validation has been extensively researched in the literature due to its importance. Our protocol follows the advanced methodologies described in previous studies, including \cite{Neigovzen2008, Feng2019, Wang2022a, Yang2022, Qu2023, Ikeda2023c}.

\subsubsection{External attack} \label{subsubsec: External Attack}

Let us now ponder on the most potent attacks that Eve has at her disposal.

\begin{enumerate}
	[ left = 0.500 cm, labelsep = 1.000 cm, start = 1 ]
	\renewcommand\labelenumi{(\textbf{A}$_\theenumi$)}
	\item	\textbf{Measure and Resend}. During the transmission of the entangled qubits, Eve intercepts them, measures them, and then returns them to their intended recipients. Eve will not find out anything by doing this because the entangled tuples do not currently carry any information. Furthermore, the decoy technique makes it easy for the legitimate players to spot a malevolent presence. The protocol is therefore totally immune to this kind of attack.
	\item	\textbf{Intercept and Resend}. Eve's plan now is to intercept the entangled particles as they are being transmitted. According to the no-cloning theorem, Eve cannot clone them. So, she prepares and sends new qubits to the designated recipients. Eve does not obtain any information by measuring them because, as we have already mentioned, they carry no information at all during the protocol's transmission part. The decoy technique also makes it possible for the legitimate players to infer that someone is listening in. The DPVQSS protocol is therefore secure against this kind of attack as well.
	\item	\textbf{Entangle and Measure}. This kind of attack starts with Eve intercepting the entangled qubits as they are being transmitted. Eve's strategy now is to entangle them with her ancilla state instead of measuring them, as opposed to the earlier attack types, and then to send them to the intended receivers. In an attempt to obtain valuable information, Eve waits for the protocol to finish before measuring her qubits. Regrettably for Eve, her actions disturb the compound system as we explain below.
	\begin{itemize}
		\item	
		In the first phase, Eve's meddling results in having $\ket{ GHZ_{ n + 2 } }$ tuples in place of the anticipated $\ket{ GHZ_{ n + 1 } }$ tuples. To obtain $\mathbf{ s }_{ i }$, for some $0 \leq i \leq n - 1$, Eve needs the $i^{ th }$ segment of the measured contents of Bob$_{ i }$'s input register, which Bob$_{ i }$ never sends over the public channel.
		\item	
		Similarly, in the second phase, Eve's tampering gives rise to $\ket{ GHZ_{ n + 2 } }$ tuples instead of $\ket{ GHZ_{ n + 1 } }$ tuples. To obtain $\mathbf{ s }$, Eve now requires the measured contents $a$ of Alice's input register, which Alice never reveals.
		\item	
		Finally, during the third phase, Eve may intercept the transmissions between Bob$_{ i }$ and Bob$_{ j }$, obtaining $\mathbf{ b }_{ i }$ and $\mathbf{ b }_{ j }$, and, ultimately, computing $\mathbf{ b }_{ j } \oplus \mathbf{ b }_{ i }$. In this manner, Eve may infer the modulo-$2$ sum $\mathbf{ s }_{ i } \oplus \mathbf{ s }_{ j }$. However, this is as far as she can get because Eve is unaware of both $\mathbf{ s }_{ i }$ and $\mathbf{ s }_{ j }$.
	\end{itemize}
	Consequently, Eve will not be able to learn anything with this strategy either.
	\item	\textbf{PNS}. The photon number splitting attack (PNS), initially described in \cite{huttner1995quantum} and further examined in \cite{lutkenhaus2000security, brassard2000limitations}, is believed to be one of the most potent strategies for an attacker like Eve to target any quantum protocol. Modern photon emitters have a recognized vulnerability: they sometimes generate multiple photons at once instead of a single one. Therefore, Eve could intercept the signals from the source intended for the distribution of the entangled particles, keep one photon from the multiple-photon signal for herself, and then transmit the rest to the legitimate receivers undetected during the transfer. However, this approach ends up being the same with the Entangle and Measure attack discussed previously. In this case, Eve again inadvertently becomes an active participant in the process, yet fails to extract any valuable information due to the reason mentioned earlier.
\end{enumerate}

\subsubsection{Inside attack} \label{subsubsec: Inside Attack}

Let us examine now what happens if there is at least on traitor among Bob$_{ 0 }$, \dots, Bob$_{ n - 1 }$. We recall that DPVQSS is a $( k, n )$ threshold QSS scheme with $k > \frac { n } { 2 }$. Moreover, we take for granted that there are at least $k$ loyal agents, i.e., the majority of Alice's agents are loyal. It is convenient to distinguish the following cases depending on the strategy a dishonest Bob may employ.

\begin{itemize}
	\item	
	As we have mentioned in subsection \ref{subsec: Verification of the Distribution Phase}, during the first phase it is possible for even a single disloyal Bob, say Bob$_{ l }$, to prevent all other Bobs from obtaining the correct information. He can do this by sending false or random data during the $C_{ 0 }^{ \downarrow }$, \dots, $C_{ n - 1 }^{ \downarrow }$ communication schemes. If Bob$_{ l }$ is rational, a standard assumption in Game Theory, he will not resort to such action because of the following reasons.
	\begin{itemize}
		\item	
		Alice will detect the presence of a traitor and she will abort the protocol.
		\item	
		Bob$_{ l }$ will miss the opportunity to obtain the whole secret.
		\item	
		Even if all traitors collaborate, they still cannot recover the secret because $k > \frac { n } { 2 }$.
	\end{itemize}
	\item	
	During the second phase, a rogue Bob$_{ l }$ may easily sabotage the verification of the initial information distribution by Alice. However, a rational Bob$_{ l }$ has no incentive to do so for exactly the same reasons we analyzed above.
	\item	
	During the third and final phase, a rogue Bob$_{ l }$ is able to sabotage the pairwise exchange of accurate information with every other Bob by sending false data. This will prevent all other agents from obtaining the real partial information vector $\mathbf{ s }_{ j }$. Unfortunately for Bob$_{ l }$, this is not enough to compromise the protocol. The way the last phase is realized, ensures that every honest Bob will truthfully and accurately relay his partial information vector, representing the piece of information entrusted to him by Alice, to every other agent, honest or not. The fact that there are at least $k$ honest agents, implies that at least $k$ out of the $n$ pieces of information will reach every loyal agent. Thus, every loyal agent will recover the secret information.
\end{itemize}

Ergo, upon successful completion of the DPVQSS protocol, all loyal agents will have obtained the secret information. As explained in Section \ref{sec: An Overview of the DPVQSS Protocol}, this criterion regarding the successful completion of the DPVQSS protocol is inspired from the quantum protocols designed for achieving Detectable Byzantine Agreement, where all loyal generals must follow the commanding general’s order. We allow for the possibility of a rogue agent also obtaining the secret information. In view of the above security analysis, we conclude that the DPVQSS protocol is correct and information-theoretically secure.

\section{Discussion and conclusions} \label{sec: Discussion and Conclusions}

This work introduces a cutting-edge $( k, n )$ threshold QSS protocol that utilizes entanglement for the establishment of a secure secret sharing scheme across a fully distributed and parallel environment. This protocol, named DPVQSS, offers the key benefits listed below.

\begin{itemize}
	\item	
	The primary aim of this protocol is to be realizable by modern quantum computers. This is achieved by employing only $\ket{ GHZ_{ r } }$ and $\ket{ \Phi^{ + } }$ entanglement. Although more intricate entangled states involving multiple particles and high-dimensional quantum states are feasible, they are challenging to create with existing quantum technology, leading to longer preparation periods and greater complexity, particularly in situations involving numerous participants. Through our method, these challenges are reduced, facilitating the practical application of the protocol.
	\item	
	Our protocol's parallel and fully distributed features are its primary innovations. This means that Alice and her agents are all spatially dispersed across various locations, making them all spatially separated. The players act on their private quantum circuits completely in parallel. Additionally, all classical communications also transpire simultaneously.
	\item	
	The DPVQSS protocol is less complicated and expensive than similar protocol because it does not rely on pre-shared keys or a quantum signature scheme. All participants use identical or similar private quantum circuits made up of Hadamard and CNOT gates, which simplifies its implementation even more and makes it practical to execute on modern quantum computers.
	\item	
	The protocol guarantees that external parties like Eve cannot obtain the secret shared by Alice, while disloyal participants themselves cannot prevent loyal agents from recovering Alice's secret.
\end{itemize}
\bibliographystyle{ieeetr}
\bibliography{DPQSSV}

\end{document}